\documentclass{aastex}
\usepackage{graphicx}
\begin{document}

\def\degrees{{^\circ }}
\def\cf{{\it cf.}} 
\def\ie{{\it i.e.}} 
\def\eg{{\it e.g.}} 
\def\vi{V-I}
\def\csq{{$\chi^2$}} 
\def\cnu{{$\chi_\nu^2$}} 
\def\spose#1{\hbox to 0pt{#1\hss}}
\def\ltsim{\mathrel{\spose{\lower 3pt\hbox{$\mathchar"218$}}
 \raise 2.0pt\hbox{$\mathchar"13C$}}}

\newbox\grsign \setbox\grsign=\hbox{$>$} \newdimen\grdimen \grdimen=\ht\grsign
\newbox\simlessbox \newbox\simgreatbox
\setbox\simgreatbox=\hbox{\raise.5ex\hbox{$>$}\llap
     {\lower.5ex\hbox{$\sim$}}}\ht1=\grdimen\dp1=0pt
\setbox\simlessbox=\hbox{\raise.5ex\hbox{$<$}\llap
     {\lower.5ex\hbox{$\sim$}}}\ht2=\grdimen\dp2=0pt
\def\gtorder{\mathrel{\copy\simgreatbox}}
\def\ltorder{\mathrel{\copy\simlessbox}}
\def\simgreat{\mathrel{\copy\simgreatbox}}
\def\simless{\mathrel{\copy\simlessbox}}

\title{A Method for Determining the Star Formation History of
a Mixed Stellar Population}

\author{Jason Harris}
\affil{Space Telescope Science Institute}
\affil{3700 San Martin Dr., Baltimore, MD, 21218}
\affil{E-Mail: jharris@stsci.edu}
\author{and}
\author{Dennis Zaritsky}
\affil{Steward Observatory}
\affil{Univ. of Arizona, Tucson, AZ, 85721}
\affil{E-Mail: dzaritsky@as.arizona.edu}

\begin{abstract}
We present a method to determine the star formation history
of a mixed stellar population from its photometry.  We
perform a chi-squared minimization between the observed photometric
distribution and a model photometric distribution, based on
theoretical isochrones.  The initial mass function, distance modulus,
interstellar reddening, binary fraction and photometric errors are
incorporated into the model, making it directly comparable to the
data.  The model is a linear combination of individual synthetic
color-magnitude diagrams (CMDs), each of which represents the
predicted photometric distribution of a stellar population of a given
age and metallicity.  While the method is similar to existing
synthetic CMD algorithms, we describe several key improvements in our
implementation.  In particular, we focus on the derivation of accurate
error estimates on the star formation history to enable comparisons
between such histories, either from different objects, or from
different regions of a single object. We present extensive tests of
the algorithm, using both simulated and actual photometric data.  
From a preliminary application of the algorithm to a subregion of the Large
Magellanic Cloud (LMC), we find that the that the lull in star
formation observed among the LMC's cluster population between 3 and 8 Gyr 
ago is also present in the field population.  The method was designed with 
flexibility and generality in mind, and we make the code available for use 
by the astronomical community.    
\end{abstract}


\section{Introduction}\label{sec:intro}
A quantitative understanding of the physical processes governing star
formation is a critical component of galaxy evolution research.  A
key step toward this understanding is the determination of detailed
star formation histories of local galaxies, in which the stellar
populations can be resolved.  The star formation history (SFH) of a
galaxy is defined by its star formation rate as a function of both age
and metallicity.  Detailed knowledge of a galaxy's SFH will enable the 
investigation of several issues in galaxy evolution, including
the chemical enrichment of galaxies, galaxy interactions as star
formation triggers, the role of stellar dynamics, and the
self-propagation of star formation.  In addition, the study of local 
galaxies provides a basic calibration for the study of more
distant, unresolved galaxies.  

The determination of SFHs from the resolved stellar photometry of
local group galaxies is a rapidly evolving area of research.  
Pioneering work in this area began by inferring a galaxy's past star 
formation rate based on the presence of stars in specific evolutionary
stages.  For example, Cepheid variables \citep{bs63, gp74} and other 
supergiants \citep{kk87} have been used to trace relatively recent
star formation activity, while carbon stars \citep{aar86, cf96}, 
asymptotic giant branch stars \citep{fb83, gal94}, RR Lyrae stars 
\citep{bs61, sms86}, and white dwarfs \citep{ns90} have been used as 
tracers of older star formation.  In contrast, classical isochrone 
fitting to a color-magnitude diagram (CMD), which is based on utilizing 
a range of stellar types, can be used to highlight the presence of a 
population of a given age and metallicity \citep[for a recent example,
see][]{mou97}.  However, this method is most useful for determining
the age and metallicity of star clusters, whose star formation
histories are simple delta functions.  One can attempt to infer the
star formation history of a galaxy from a set of such cluster
measurements \citep{gir95, oz96}, but it is unclear how the cluster
formation rate and the overall star formation rate are related.  

These methods provide a crude picture of a galaxy's SFH, but
they cannot be used to quantitatively determine the relative
numbers of stars formed as a function of age and metallicity in a
mixed population.  The methods developed to provide such a
quantitative measure of the SFH of mixed stellar populations fall
broadly under the category of synthetic CMD fitting.  A synthetic CMD
is the expected photometric distribution of stars, given a particular
SFH, photometric errors, and values for the distance modulus, initial
mass function (IMF) slope, binary fraction, and interstellar
extinction.  All synthetic CMD methods involve the construction of a
quantitative fitting statistic, with which synthetic CMDs are compared
to an observed CMD.  A minimization algorithm is then used to 
determine the SFH of the observed population by constructing a
synthetic CMD that most closely resembles the observed CMD.  Some
synthetic CMD methods use the main sequence (MS) luminosity function
(LF) to construct the fitting statistic \citep{tos91, sta97, hol97,
ard97, alo99}, while others use the LF of particular post-MS phases of
stellar evolution.  For example, \citeauthor{dp97} used the LF of blue
supergiant Helium-burning stars to reconstruct the recent SFH of
Sextans A \citep{dp97} and GR8 \citep{dp98}.   

The inclusion of the available color information into a synthetic 
CMD fitting method began with the R-method \citep{ber92, val96a,
val96b, geh98}, in which the ratios of the numbers of stars in
photometrically-defined evolutionary states (\eg, the ratio of the
number of MS stars to red giants) is combined with the LF to determine
the SFH.  The R-method is the direct precursor to the most recent
synthetic CMD fitting methods, in which most (if not 
all) of the two-dimensional information of the CMD is used to
quantitatively discriminate between model SFHs.  In these methods, the
CMD plane is divided into subregions, and the number of stars present
in each subregion of both the observed and synthetic CMDs is recorded.
These subregion populations are used to construct the fitting
statistic, and determine the best-fit SFH.  Detailed descriptions of
synthetic CMD methods are given in \cite{ts96}, \cite{apa96},
\cite{dol97}, and \cite{her99}.  Examples of synthetic CMD methods
applied to local group galaxies include NGC 6822 \citep{gal96a,
gal96b}, NGC 185 \citep{md99}, the Pegasus dwarf irregular galaxy
\citep{apa97a, ggr98}, the Carina dwarf irregular galaxy \citep{hk98},
Leo A \citep{tol98}, Leo I \citep{gal99}, LGS3 \citep{apa97b}, and the
Large Magellanic Cloud \citep{hol99, ols99}. 

In this paper we present our own synthetic CMD-based SFH algorithm.
Our algorithm is similar to the methods listed above, with what we
consider to be a few key improvements: 
(1) the algorithm efficiently searches large parameter spaces, making 
artificial constraints on the SFH (\eg\ an imposed chemical 
enrichment law) unnecessary;
(2) it takes full advantage of all available photometric
information (the algorithm uses multiple CMDs, and all photometered
stars can contribute to the fit); 
(3) it allows for a more realistic treatment of interstellar
extinction that includes both differential extinction and age
dependence;  
(4) it performs a detailed calculation of the uncertainties in the 
best-fit SFH, based on the \csq\ topology of the parameter space
surrounding the best-fit minimum; and 
(5) it is designed with generality and flexibility in mind,
so that members of the astronomical community can apply it to their
own data, using any set of isochrones.

While the generality and flexibility of the method are important, we
present the method in a slightly specialized form in this paper,
as a primer for our upcoming analysis of the SFH of the Magellanic
Clouds (MC).  Since our MC data consist of $UBVI$ photometry, we
construct synthetic CMD triplets ($\ub$ vs. $B$; $\bv$ vs. $V$; $\vi$
vs. $I$) for model/data comparison.  Unless otherwise noted, our
examples also use the distance modulus, IMF, interstellar extinction,
binary fraction, and photometric errors appropriate for these data.  

We provide a brief overview of the method in \S\ref{sec:method}.  In
\S\ref{sec:basis}, we describe the construction of synthetic CMDs from
theoretical isochrones.  The minimization algorithm is described in
\S\ref{sec:min}.  We test the method using both artificial photometry
and real data in \S\ref{sec:tests}.  We summarize the results in
\S\ref{sec:summ}. 

\section{Method Description}\label{sec:method}

\begin{figure}
\begin{center}
\includegraphics{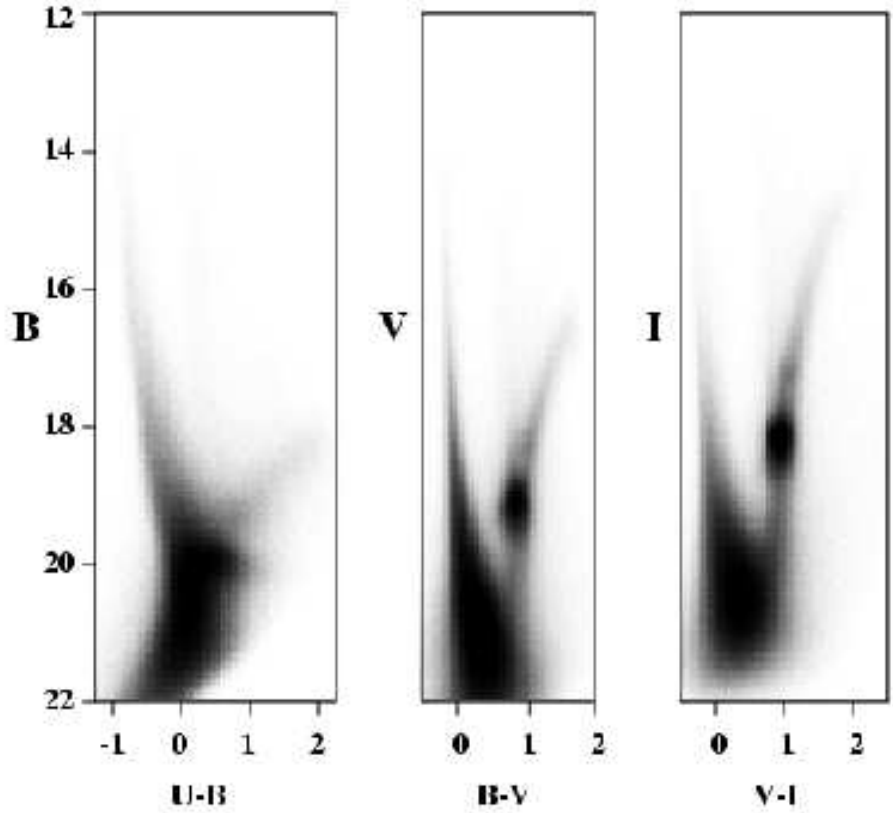}
\end{center}
\caption{Three projections of the 4-dimensional photometric
distribution of 4.1 million stars from our MC Survey.  From left to
right: the $\ub$ vs. $B$, $\bv$ vs. $V$, and $\vi$ vs. $I$
color-magnitude diagrams.  The star formation history of these stars
is encoded in their current photometric distribution. \label{fig:ubvi}}
\end{figure}

A CMD triplet representing our $UBVI$ photometry of 4.1 million stars
from a $4\degrees \times 2.6\degrees$ region of the LMC is shown in
Figure \ref{fig:ubvi} (\cf\ \cite{zht97} for a detailed description of
the data).  Other than the star formation history itself, several
external astrophysical parameters influence the observed photometric
distribution: the distance of the stellar population, the
interstellar extinction, the initial mass function (IMF), the binary
fraction, the line-of-sight depth of the population, and the
photometric errors.  Each of these factors can be independently
constrained, leaving only the SFH to be determined. 

We begin the SFH reconstruction with a library of theoretical
isochrones.  Each isochrone describes the intrinsic photometry of
a stellar population with a particular age and
metallicity.  We transform that photometry into a synthetic CMD by
accounting for each of the parameters listed above.  We consider each
synthetic CMD to be an {\it eigen-population}: the photometric
distribution that would be observed for a stellar population with the
age and metallicity of the parent isochrone.  While the
{\it eigen-populations} are not orthogonal, we take steps to ensure
that each synthetic CMD represents a unique photometric distribution
(\cf\ \S\ref{sec:synth}).

A linear combination of these {\it eigen-populations} forms a
composite model CMD that can represent any SFH.
The amplitude associated with each {\it eigen-population} is
proportional to the number of stars formed at that age and
metallicity.  The best-fit SFH is described by the set of amplitudes
that produces a composite model CMD most similar to the
observed photometry.  The best-fitting amplitudes are determined
using a modified downhill simplex algorithm, described in detail in
\S\ref{sec:min}.  

A key advantage of this {\it eigen-population} approach is that
once the library of synthetic CMDs is constructed, we can construct 
composite model CMDs and compare them to the observed photometry with 
little computational cost.  This efficiency allows us to explore
large parameter spaces; we have employed up to 50 independent SFH
amplitudes.  The efficiency comes at the cost of some flexibility; the 
external parameters discussed previously (the distance modulus,
extinction, IMF, binary fraction, line-of-sight depth, and
photometric errors) are incorporated into the synthetic CMDs, and so
it is not possible to solve for these parameters and the SFH
amplitudes simultaneously.  Values for these parameters must be  
determined (or assumed) prior to the SFH analysis.  However, the SFH
analysis can be repeated with different parameter values, and the
results compared.  In this way, we explore the sensitivity of the
fit to these external parameters (\cf\ \S\ref{sec:tpvary}).

\section{Constructing the Synthetic CMDs}\label{sec:basis}

\subsection{Theoretical Isochrones}\label{sec:isoc}

\begin{figure}
\begin{center}
\includegraphics{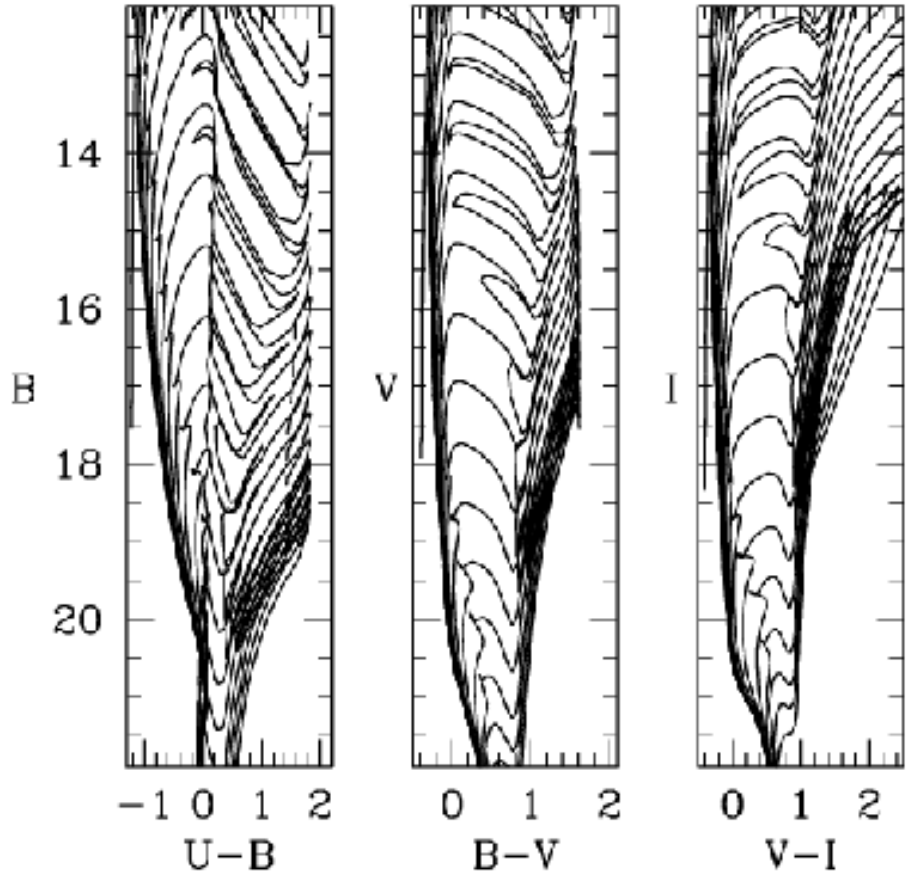}
\end{center}
\figcaption{A sample of the Padua Isochrones \citep{ber94}.  Shown
are the z=0.008 isochrones, with ages ranging from 4 Myr to 16 Gyr,
with logarithmic time steps of 0.2.  The isochrones are shown with a
distance modulus of 18.5, and with the same photometric limits as
Figure \ref{fig:ubvi}. \label{fig:isoc}}
\end{figure}

\begin{figure}
\begin{center}
\includegraphics{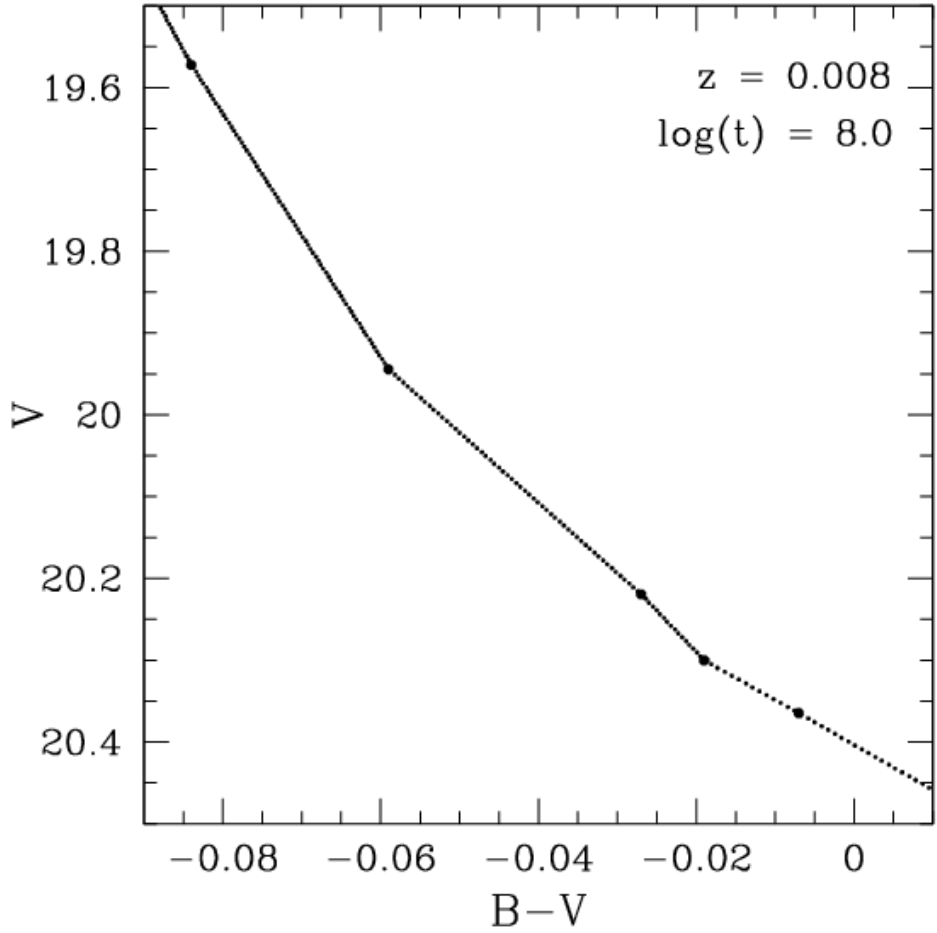}
\end{center}
\figcaption{Interpolating the isochrones.  Along the main sequence,
the Padua isochrones are sparsely populated.  The large dots are the
original Padua isochrone points; the small dots are our interpolated
points. \label{fig:interp}}
\end{figure}

\begin{figure}
\begin{center}
\includegraphics{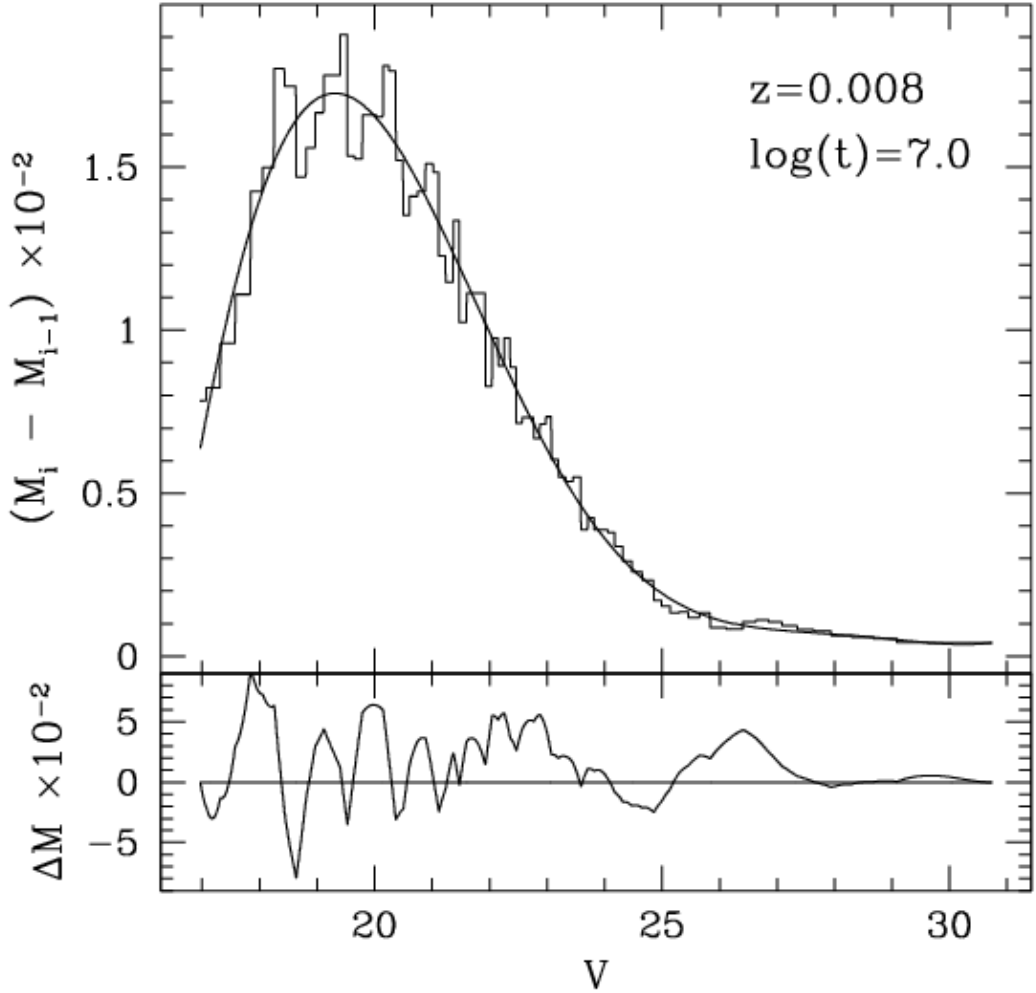}
\end{center}
\figcaption{The mass difference between adjacent main sequence
isochrone points as a function of $V$ magnitude, for a typical
interpolated isochrone (thin line).  These mass differences determine
the occupation probability of each location along the isochrone.
They should be smoothly distributed, because observed
luminosity functions of main sequence stars are smooth.  The lumpiness
is an artifact caused by the fact that the isochrone points follow
straight line segments in the CMDs (Figure \ref{fig:interp}).  We
correct for this artifact by fitting a smooth curve through the mass
differences (heavy line in the figure), and using the mass implied by
the curve for each isochrone point. \label{fig:dmass}}
\end{figure}

\begin{figure}
\begin{center}
\includegraphics{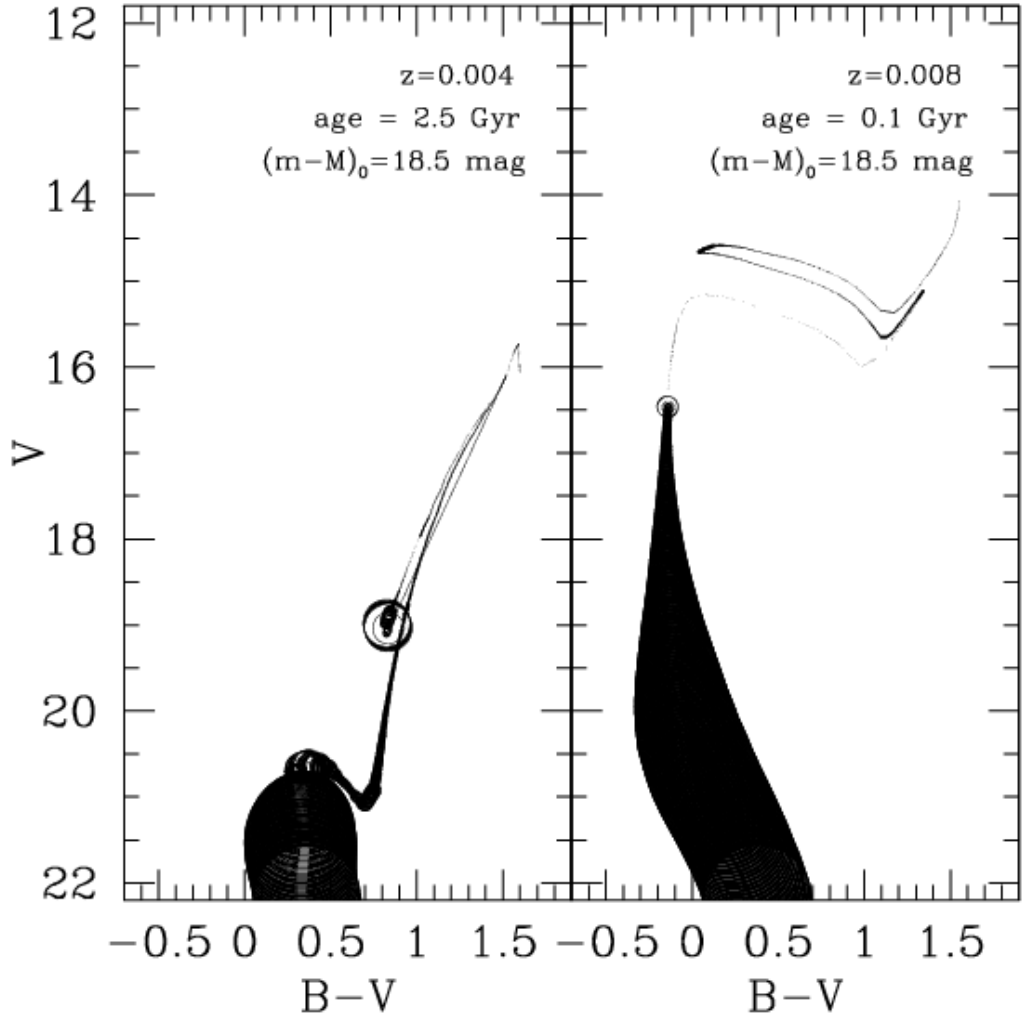}
\end{center}
\figcaption{Two sample isochrones plotted with occupation probability
represented by point size. \label{fig:op}}
\end{figure}

Reconstructing the SFH of a stellar population from its photometry
begins with theoretical isochrones.  We use the Padua set of
isochrones \citep{ber94, gir00}, because it covers a wide range of
ages and metallicities, and includes He-burning phases of stellar
evolution (\ie, the red clump).  A representative sampling of the
isochrone photometry is shown in Figure \ref{fig:isoc}.  We find that
the isochrones are too coarsely sampled along the main sequence for
our purposes, so we perform a linear interpolation of the photometry.
The interpolated points are evenly distributed along the isochrone
with a separation much smaller than the characteristic photometric
errors ($\Delta m=0.005$ mag, \cf\ Figure \ref{fig:interp}).  This
default level of interpolation will work well for any currently
available data, but like most of the algorithm's parameters, it can
easily be changed if necessary.  We then assign to each isochrone
point a relative  occupation probability (OP) based on an assumed IMF
slope: 

$$OP_i = C \times (M_i^\Gamma - M_{i-1}^\Gamma)$$

\noindent where $M_i$ and $M_{i-1}$ are the masses of adjacent
isochrone points and $\Gamma$ is the logarithmic IMF slope (unless
otherwise noted, we adopt the Salpeter IMF; $\Gamma= -1.35$).  $C$ is
a normalization constant, with a value chosen such that the total
integrated probability over $0.1 < M/M_{\odot} < 100.0$ is unity.
The highly magnified view of the isochrone in Figure \ref{fig:interp}
reveals that the interpolated points lie along straight line segments
in the CMD plane, in approximation of the true isochrone curve.
The interpolated masses have similar approximation errors, as 
illustrated in Figure \ref{fig:dmass}, where we plot the mass
differences between adjacent isochrone points as a function of $V$
magnitude.  The distribution of approximated mass differences is not
smooth, and this lumpiness propagates to the OP values.  To remedy
this artificial lumpiness, we fit a smooth polynomial through the
points of Figure \ref{fig:dmass}, and adopt the mass implied by the curve
for each isochrone point.  We vary the order of the fitted polynoial
from 1 to 12, and determine the best-fit polynomial in each case.  Of
these candidates, the overall best-fit polynomial is the one with the
lowest \csq\ value.  The fits are examined interactively to ensure that
the data are not overconstrained.  The resulting mass correction is
never more than several hundreths of a solar mass.  We smooth only
those isochrone points that are fainter than the main sequence
turn-off (MSTO).  Above the MSTO, the original isochrone points are 
sufficiently dense that our interpolated points do not deviate
significantly from the true isochrone curve.  The final distribution
of OP values is illustrated in Figure \ref{fig:op}, in which the
relative OP is represented by point size for two sample isochrones. 

\subsection{Interstellar Extinction}\label{sec:redden}

In previous work, interstellar extinction has typically been modeled
as a single mean value, which is simple to implement but implies an 
unphysical foreground-sheet dust geometry.  \cite{hzt97} investigated
the distribution of line-of-sight reddening values toward over 2000 OB
stars in a small region of the LMC and found that the distribution of
extinction values is much wider than can be accounted for by the
photometric errors.  Rather, the distribution is consistent with
clumpy, exponential line-of-sight distributions for both OB stars and
dust, with the scale height of OB stars roughly half that of the dust.

Further evidence of differential extinction in the LMC was presented
by \cite{zar99}.  \citeauthor{zar99} demonstrated that extinction in
the LMC is population-dependent; hot stars have an average $A_V$
extinction that is four times larger than that of cool stars (\cf\ 
Zaritsky's Figure 12).  The population dependence of the
extinction can be understood as a stellar dynamical effect.  The hot
stars come from recently formed stellar populations, which tend to be
located in regions of the galaxy where the gas and dust density is
high.  The cool stars are predominantly red giants, and are at least
a few billion years old.  These stars have dispersed from their dusty
formation sites and therefore tend to lie along lines of sight with
lower extinction.  

For our LMC analysis, the synthetic CMDs incorporate an age-dependent 
extinction: stars from young isochrones ($\log(t)<7.0$) are assigned
extinction values drawn randomly from Zaritsky's hot star sample, and
stars in old isochrones ($\log(t)>9.0$) are assigned extinctions drawn
from Zaritsky's cool star sample.  Stars of intermediate age can get
their extinction value from either sample, with a probability function
that varies linearly with $\log(t)$, increasingly favoring the cool
star sample as the isochrone age increases.  The age divisions are
somewhat arbitrary; $\log(t)=7.0$ is roughly the main sequence
lifetime of OB stars, and $\log(t)=9.0$ is an approximation of the
diffusion timescale of unbound clusters in the LMC
\citep[\cf][]{hz99}.  This reddening prescription was developed for
use on our MC Survey photometry; the reddening prescription can 
easily be adjusted by the user to be appropriate for any input data. 

\subsection{Photometric Errors}\label{sec:errs}

Artificial star tests (ASTs) provide an accurate estimate of the
photometric errors, as a function of magnitude and color
\citep[\cf][]{sh88}.  A set of artificial stars with known photometry
is inserted into the frame, and the frame is analyzed using the
standard data reduction pipeline to recover the photometry of all
stars, including the artificial stars.  

In performing artificial star tests, the artificial stars
must not significantly overlap, or the frame's crowding conditions
will not be correctly sampled.  To this end, we follow the procedure
outlined by \cite{gal99}.  We add artificial stars to the frames in a
regular grid, separated by 12 pixels (8.4 arcsec) in each direction.
Because the median seeing of the MC Survey photometry is 1.4 arcsec,
the artificial stars will overlap only beyond their
$3\sigma$ radii.  Positions for the artificial stars are generated in 
right ascension and declination, and for each of the $UBVI$ images, we
invert the coordinate solution to determine the corresponding pixel
coordinates.  This procedure ensures that the artificial stars are
added to the image for each filter at the same location with respect
to the real stars, regardless of any pointing offsets between
exposures.   

To obtain a statistically robust determination of the photometric
errors, the total number of artificial stars added should be 
significantly larger than the number of real stars detected in the
image.  Since a single artificial star test typically contains no
more than 5-10\% of the image's total star count, several dozen tests
must be run on each image.  We randomly offset the grid of artificial
stellar positions between runs, so that the artificial stars in each
run sample unique crowding environments. 

It is also necessary to match the frame's distribution of stellar
magnitudes when constructing the artificial star list, because the
photometric errors are dependent on the brightnesses and colors of the
stellar sample.  We construct the artificial photometry by randomly
selecting stars from the theoretical isochrones, according to the
Salpeter IMF.  In addition, we simulate differential extinction for
young stellar populations by assigning extinction values randomly
between $A_V=0.0$ mag and $A_V=1.5$ mag for stars younger than 100
Myr.  Stars older than 100 Myr are assigned $A_V=0.0$ mag.  These
steps ensure that the distribution of artificial magnitudes grossly
matches that of the data frames, for each of our $UBVI$ filters. 

\begin{figure}
\begin{center}
\includegraphics{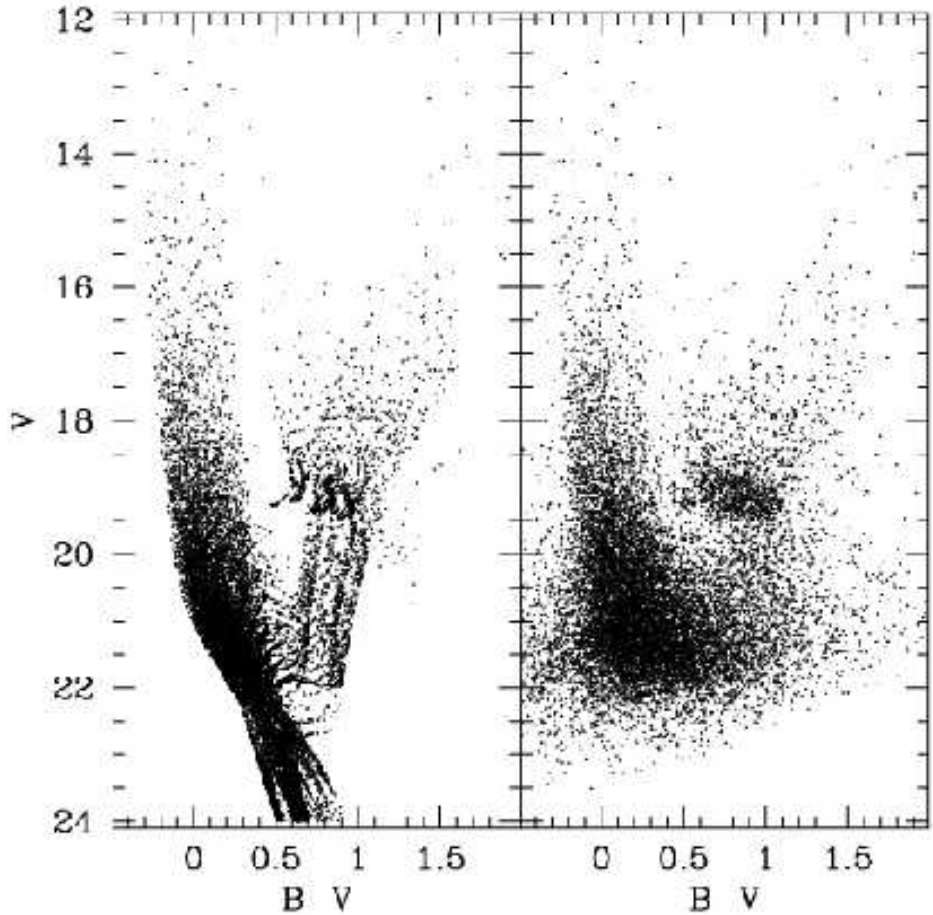}
\end{center}
\figcaption{{\bf Left:} Sample input photometry for an
artificial star test.  The photometry is drawn from isochrones,
assuming a distance modulus of 18.5 mag, a generic SFH, and a Salpeter
IMF.  In addition, young isochrones are differentially reddened.  Note
that the main sequences are populated well below the expected faint
limit, in order to calculate completeness fractions.  {\bf Right:} The
photometry as recovered by our artificial star tests.  The tests
recover artificial stars from all four of our $UBVI$ frames; only the
$\bv$ vs. $V$ CMD is shown here. \label{fig:ast}}
\end{figure}

Having determined the input coordinates and $UBVI$ magnitudes for the 
artificial stars, we add them to the images using the DAOPHOT ADDSTAR 
routine.  The resulting images are then processed using
the same automated pipeline that we use on the original data.  The
pipeline applies the coordinate solutions and matches up the $UBVI$
frames to construct a $UBVI$ photometric catalog, containing both
artificial and real stars (the combined output list).  To extract the 
recovered artificial star photometry from the combined output list,
we use the input catalog of artificial stars and the original stellar
catalog containing only real stars.  We begin by removing stars from 
the combined catalog whose coordinates are matched to within 1 arcsec
by a star in the original catalog.  To avoid accidentally removing
artificial stars from the combined list, we further require that the
matched stars have similar photometry (within 0.5 mag in each filter)
in both lists.  Stars in the trimmed output list are then matched with
the input artificial stars list, so that we have both the input and
recovered $UBVI$ magnitudes for each artificial star.  If no match for
an artificial star is found in the trimmed output list, that star is 
flagged as a photometric dropout for calculation of the completeness
rate.  The input and recovered photometry from a typical artificial
stars test is shown in Figure \ref{fig:ast}.

To determine the dependence of the photometric errors on brightness
and color, we divide the artificial star CMDs into magnitude and color
bins, and construct the $\Delta m = m_{recovered} - m_{inserted}$
distribution for each bin.  One could assign 
photometric scatter to model stars by identifying the CMD bin in which
the model star is found, and drawing randomly from that bin's $\Delta
m$ distribution.  However, this procedure results in discontinuities
in the photometric scatter at bin boundaries.  To ensure that the
photometric scatter is smooth throughout the synthetic CMDs, we
interpolate between the $\Delta m$ distributions of the four CMD bins
surrounding each model star.  

\subsection{The Synthetic CMDs}\label{sec:synth}

Once we have the library of isochrones, the reddening statistics, and
the artificial star test results, we construct synthetic CMDs to be
compared to the photometric data.  Our isochrone library contains 142
isochrones spanning ages from 4 Myr to 18 Gyr, and metallicities
between z=0.001 and z=0.008.  We decided not to construct an
independent synthetic CMD for each of these isochrones, because adjacent 
isochrones differ in age by only $\Delta \log(t)=0.05$ years, and are
often photometrically nearly degenerate.  We therefore combine
adjacent isochrones into age groups covering $\Delta \log(t)=0.2$
years, reducing the number of independent model parameters to 40. 
The need to combine isochrones into groups is dictated by the quality
of the input data.  Data with much smaller photometric errors at faint
magnitudes will not require such a constraint (\cf\ \S\ref{sec:gc}).

We construct one synthetic CMD triplet ($\ub$ vs. $B$, $\bv$ vs. $V$,
and $\vi$ vs. $I$) for each isochrone group.  By selecting one million
random points from the isochrones' OP distributions, we construct the
predicted photometry for an extremely large stellar population formed
at the age and metallicity of the parent isochrones.  We construct a
large stellar population for each synthetic CMD to ensure that 
the Poisson noise associated with the model is small compared to the
Poisson noise of the data.  We apply a distance modulus of 18.5 mag to
each of the star's magnitudes, placing it at the distance of the LMC. 
In general, the distance modulus applied to the model stars should
vary according to a model of the real population's line-of-sight 
distribution.  However, the line-of-sight depths of the LMC and of the
globular clusters we examine in this paper are insignificant, so we
apply a single distance modulus in the present work.  The model stars
are reddened by drawing a random extinction value from the appropriate
stellar sample in \cite{zar99} (\cf\ \S\ref{sec:redden}).  We then
apply the photometric error model derived from a typical LMC
artificial stars test (\cf\ \S\ref{sec:errs}).  For each model star,
we interpolate the completeness fraction from the surrounding CMD
bins, and probabilistically remove the star from the population
according to this completeness fraction.  If the star is not removed,
its photometry is scattered by randomly drawing from the interpolated 
$\Delta m$ distributions in each CMD dimension.  

We bin each synthetic CMD into 0.05 mag $\times$ 0.05 mag pixels.  The
pixel values are computed as: $p(i) = N(i)*f/N$, where $p(i)$ is the
value of pixel $i$, $N(i)$ is the number of model stars observed in
pixel $i$, $N$ is the total number of stars in the model population,
and $f$ is the fraction of stars represented in the current isochrone:

$$f = \frac{M_{max}^\Gamma - M_{min}^\Gamma}{100.0^\Gamma -
0.1^\Gamma}$$

\noindent Each isochrone covers a finite range in masses that is a
subset of the full mass interval over which stars form.  Stars outside
the mass range covered by the isochrone have either evolved to
non-luminous end states, or are less massive than the smallest mass
represented in the Padua isochrones, $M_{min}=0.6 M_\sun$. After this 
normalization, the pixel values represent the fraction of all stars
formed in the synthetic population that currently occupy each CMD
pixel.  This normalization is critical, as it provides a direct
transformation from the number of stars observed in a CMD to the
number of stars formed in a population.  

\begin{figure}
\begin{center}
\includegraphics{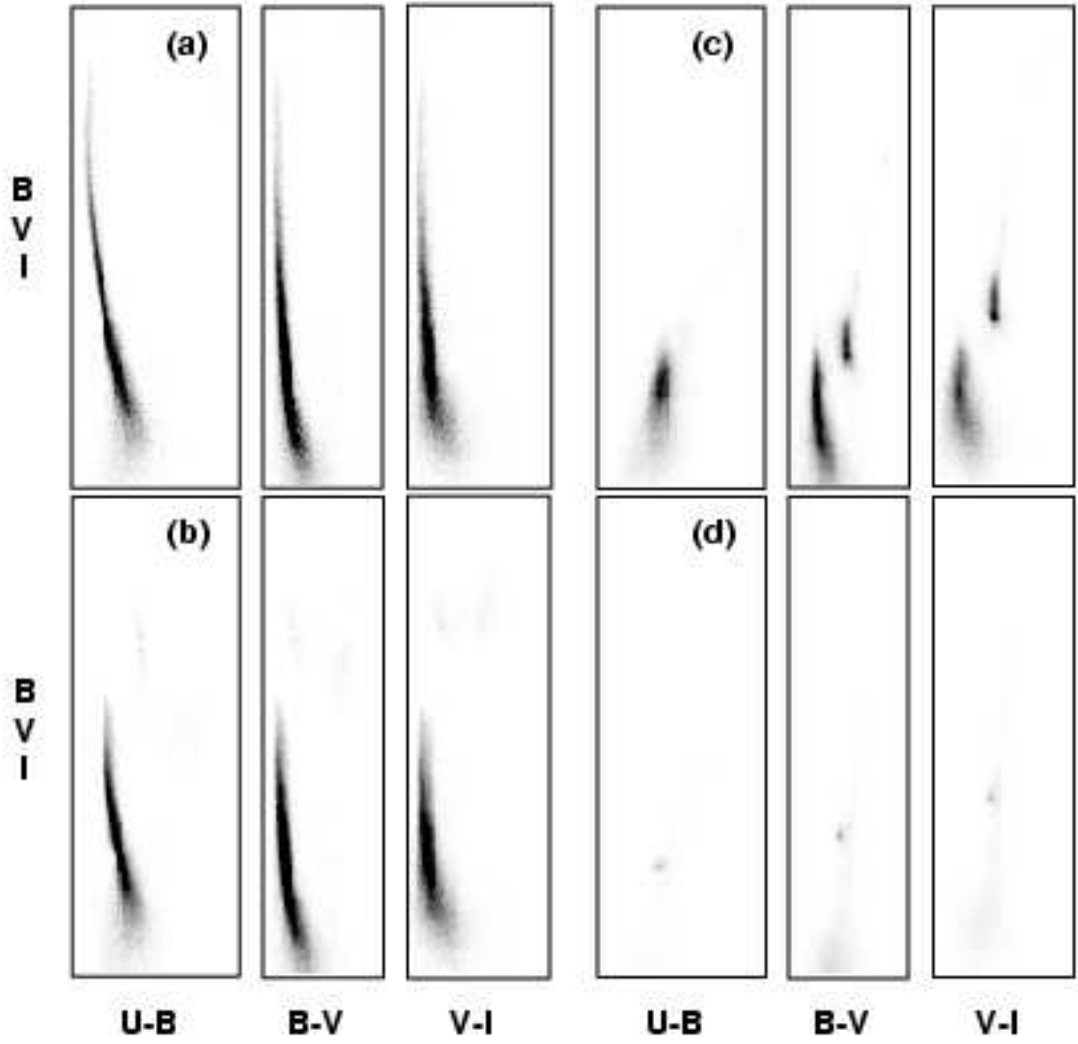}
\end{center}
\figcaption{A sampling of our synthetic CMD library.  The limits on
each CMD are as in Figure \ref{fig:ubvi}.  Four CMD triplets are
shown: 
{\bf (a)} z=0.008, age=$10^7$ yr; {\bf (b)} z=0.008, age=$10^8$ yr;
{\bf (c)} z=0.004, age=$10^9$ yr; {\bf (d)} z=0.001, age=$10^{10}$ yr;
\label{fig:synth}}
\end{figure}

This procedure results in a library of synthetic CMD triplets, one
triplet for each isochrone group.  A representative sample of the
synthetic CMDs is shown in Figure \ref{fig:synth}.  Pixel values in
the older synthetic CMDs are smaller, illustrating the
normalization outlined above.  A composite model CMD is constructed as
a linear sum of the synthetic CMDs.  Each term in the sum is assigned
an amplitude coefficient that represents the number of stars formed at
the age and metallicity of that synthetic CMD.  We impose a positivity
constraint on these amplitudes to avoid unphysical negative star
formation rates.  By varying the amplitudes, we are able to construct
a model CMD that represents the observed photometry of a stellar
population with any arbitrary SFH. 

\section{The Minimization Algorithm}\label{sec:min}

The amplitude coefficients modulating the synthetic CMDs form a 
N-dimensional parameter space of possible SFHs.  Given an observed CMD
triplet, we assign a fitting statistic (\csq) to
each set of amplitudes that describes how well that model's 
CMDs match the observed CMDs.  The best-fitting SFH is then
described by the set of amplitudes with the lowest \csq\ value.
We construct \csq\ as the sum:

$$\chi^2 = \sum_{i} \frac{(N_{D_i} - N_{M_i})^2}{N_{D_i}}$$

\noindent where $N_{D_i}$ is the number of stars observed within
the $i$th subregion of the CMD triplet, and $N_{M_i}$ is the
number of stars in the composite model, in the same subregion.  The
division of the CMDs into subregions is specified by the
user.  We have experimented both with a uniform gridding of the CMD
planes into boxes covering $0.25 \times 0.25$ mag, and with an
adaptive grid that is fine in photometrically dense regions like the
red clump and main sequence, and coarse in photometrically sparse
regions.  We found the best-fit SFH to be relatively insensitive to our
choice of gridding strategy; because it is simpler, we use the uniform
gridding in the present work.  The grid size of 0.25 mag was chosen as
detailed enough that we can discriminate among similar SFH models, but
coarse enough that we will not need to interpolate between isochrones.
Subregions in which $N_{D_i} = N_{M_i} = 0$ do not contibute to the
sum.  If $N_{D_i} = 0$, but $N_{M_i} > 0$, we replace $N_{D_i}$ with
$1.0$ in the denominator, as an approximation of the Poisson error
when $N=0$.  

Previous SFH studies have had sufficiently few parameters that a brute
force minimization algorithm, in which all of parameter space is
tested, was practical.  However, we are constructing parameter spaces
with up to 51 dimensions (51 independent isochrone amplitudes), so
searching all of parameter space is not possible.  Instead, we employ
a downhill simplex \csq\ minimization algorithm \citep{nrf92}.  We
start at a random point in parameter space, and construct a {\it
simplex} about that point.  The simplex is a collection of $N_{dim}+1$
parameter locations and their corresponding \csq\ values, where
$N_{dim}$ is the number of parameter space dimensions.  The first
point is the original random location, and the remaining points are
generated by perturbing the original point by a small fixed amount
along each dimension in the parameter space.  Thus, a
three-dimensional simplex would be a tetrahedron.  From the collection  
of \csq\ values, the local \csq\ gradient is calculated, and a step in
the direction of the gradient is taken.  The new local \csq\ gradient
is calculated, and the process repeats until a minimum is found.  

We use two safeguards against settling on local, rather than global,
minima.  First, when the simplex signals that it has found a minimum,
it is expanded in all dimensions, and allowed to reconverge.  Once
this reconvergence returns the simplex to the original parameter
location, our second safeguard is activated.  A small step is taken
in a random parameter space direction.  If the \csq\ value of the new
location is lower than the minimum candidate, then the candidate was a
local minimum.  We continue stepping in this direction until an
``uphill'' step is taken. This is repeated for tens of thousands of
random parameter space directions, and the simplex is restarted at the
tested location with the lowest \csq\ value. 

We find that this second safeguard is critical to finding the true
global minimum.  On its own, the simplex often converges on local
minima because it only searches the local parameter space along 
directions parallel to the $N_{dim}$ parameter axes.  As the simplex
approaches the minimum, it becomes more important to explore
parameter changes where a deviation in one parameter is offset by a
complementary deviation in one or more other parameters.  In other
words, downhill gradients tend to lie along ``off-axis'' parameter
directions near the minimum, but these directions are ignored by the
simplex.  Therefore, we iterate between converging toward a minimum
with the simplex, and searching in random parameter directions
for lower \csq\ values.  The iteration stops only when the search of
random directions produces no lower \csq\ values.  The number of
random directions searched is arbitrary, but our choice of $\sim30000$
directions appears to cover the parameter space sufficiently.  Tests
in which we increase the number of search directions by a factor of 10
produce the same global minima.  

An important issue in such \csq\ minimization procedures is whether
the best-fit model is a good absolute fit.  The canonical determinant
is the \csq\ value itself: the best-fit reduced \cnu\ should be
roughly unity for a good fit.  However, this criterion is rigorously
true only if the errors on the fit are Gaussian distributed, which is
unlikely in this  application.  Nevertheless, we can empirically
calibrate \cnu\ by determining the best-fit SFH of model input
photometry, constructed from the same isochrones, reddening statistics
and crowding effects used to construct the synthetic CMDs.  The \cnu\ 
values of these artificial data provide an expectation for models that
match the data accurately. Despite the non-Gaussian errors, we find
that $\chi_\nu^2 \sim 1.0$ is still indicative of a good fit to the
data (\cf\ \S\ref{sec:tests}).  

Error bars on the SFH amplitudes are determined by identifying the
68\% ($1\sigma$) confidence interval on each amplitude.  The
confidence intervals are determined from the \csq\ topology of the 
parameter space near the minimum.  We explore this topology by
sampling the \csq\ values of selected points near the minimum.  These
test points are selected in three different ways:
First, we vary each amplitude in turn, while holding all others fixed,
to determine the uncorrelated error on each amplitude.  
Next, we vary adjacent pairs of amplitudes while holding others fixed,
to determine the pairwise correlated errors.  
Finally, we vary all amplitudes simultaneously, to determine the
general correlated errors.  Although the first two strategies are
subsets of the more general third strategy, we employ all three
strategies to improve the efficiency of the search.  Exploring the
\csq\ topology of a 40-dimensional space is a daunting task; searching
along hundreds of thousands of random directions does not necessarily
cover the local topology with sufficient density to construct reliable
error bars.  We employ the uncorrelated errors and
pairwise correlated errors to efficiently increase the sampling
density in these important cases.  In each case, we
iteratively step away from the minimum, and determine the \csq\ value
until \csq\ increases to the 68\% confidence limit.  From the
collection of tested points, we select the largest and smallest
values for each amplitude as the limits for the errorbars.

\section{Testing the Method}\label{sec:tests}

We test our SFH minimization algorithm by recovering the SFH of both
artificially-generated photometry and real data.  The artificial
photometry is based on an input SFH, the theoretical isochrones, and
reddening and crowding statistics derived from real data.  The
SFH recovery of these artificial populations allows us to test the
performance of our algorithm, and its dependence on parameters such as
the reddening distribution, crowding statistics, and the IMF slope.  Since 
we can construct artificial populations with complex SFHs, these tests also
illustrate the algorithm's ability to disentangle mixed stellar populations, 
and recover the correct SFH.  For the real photometry, we use deep globular 
cluster photometry and selected regions of our Magellanic Clouds survey.  
The globular cluster data are used to ensure that no systematic bias is 
introduced by our synthetic-CMD-fitting method, compared to more 
traditional isochrone-fitting techniques.

\subsection{Artificial Photometry Tests}\label{sec:artphot}

We recover the SFH of artificial stellar populations in the manner 
described in \S\ref{sec:min}.  First, we construct artificial stellar
populations using the same values for external parameters (IMF slope,
distance modulus, extinction, and binary fraction) as we used to
construct the synthetic CMDs.  By holding the parameter values fixed,
we will highlight any systematic problems with our algorithm, and 
illustrate the goodness of fit that can be expected from the
algorithm, if all the parameters are correctly determined.  We then
vary the external parameter values in constructing the artificial
populations, but we continue to use synthetic CMDs based on the
original parameter values.  In this way, we examine the effect
on the recovered SFH of having incorrectly determined these external
parameters. 

\subsubsection{Parameter-matched populations}\label{sec:tpmatch}

Artificial populations constructed using the same external parameter 
values as adopted in the fitting algorithm allow us to examine the
performance of the algorithm directly.  In each of these tests, we use
canonical values appropriate for the LMC: $(m-M)_0=18.5$ mag,
$\Gamma=-1.35$, $f_{binary}=0.5$, the population-dependent extinction
distribution described in \S\ref{sec:redden}, and the photometric
errors derived from a typical artificial star test from our MC survey
data.  

\begin{figure}
\begin{center}
\includegraphics{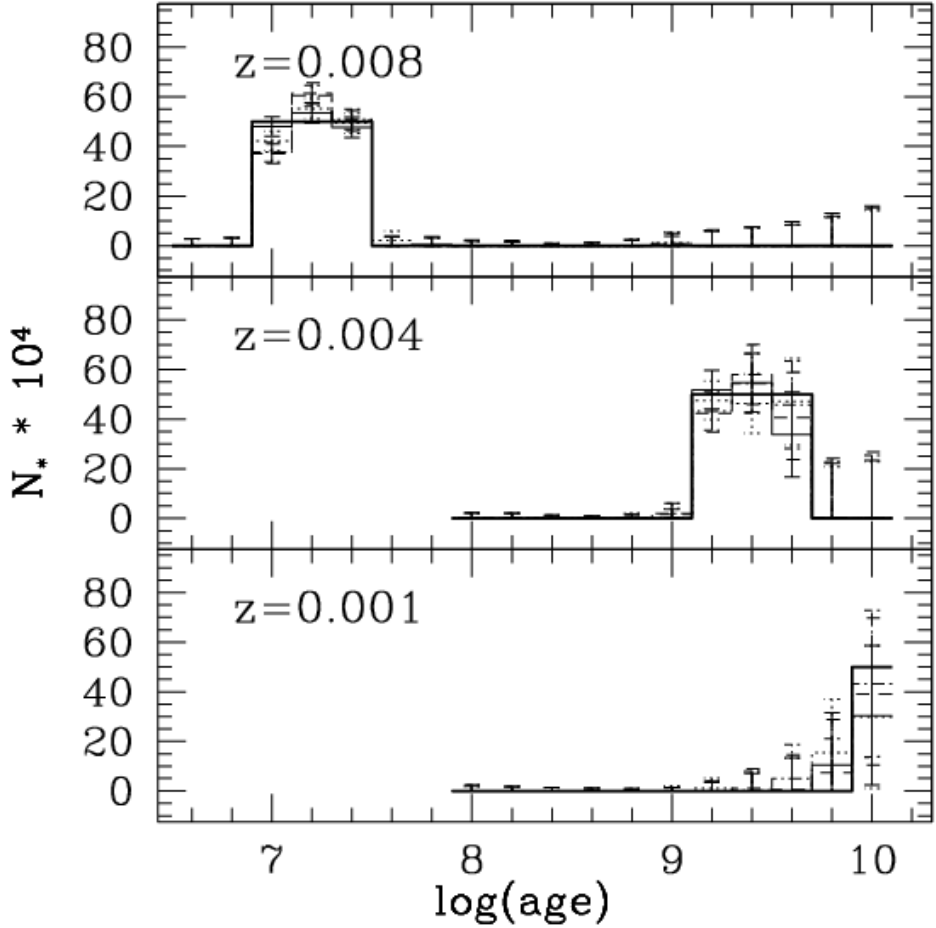}
\end{center}
\figcaption{The recovery of a simple, extended-burst SFH.  The
histograms indicate the number of stars formed in each isochrone
group.  The isochrones have been divided according to their
metallicity.  The top panel shows isochrones with z=0.008; the middle
panel shows isochrones with z=0.004; and the bottom panel shows
isochrones with z=0.001.  The input SFH is shown as a heavy solid
line.  It consists of a young, metal-rich burst, an old metal-poor 
burst, and an intermediate burst.  Four artificial stellar populations 
are generated from this SFH; the random seed used to select the
photometry of the stars is varied in each instance.  The best-fit
recovered SFH for each of these populations is shown as a dotted line.
The errorbars indicate the 68\% confidence interval on each amplitude
in the \csq\ fit. 
\label{fig:test1}}
\end{figure}

The first populations are constructed from a simple SFH in which there 
are three extended bursts of star formation, ranging from old and 
metal-poor to young and metal-rich.  The star formation rate (SFR) is
held constant for the duration of each burst.  We generate four
populations from this SFH, varying both the random seed used to select
points along the isochrones, and the random seed that selects the
starting set of amplitude values. The recovered SFHs of these
populations are shown in Figure \ref{fig:test1}.  We conclude the
following from these tests:  
(1) we have successfully recovered the SFH of the input population,
(2) our errorbar estimates are reasonable because they are
approximately equivalent to the variance of the amplitudes when
different random seeds are used to construct the artificial
population, and 
(3) starting the simplex at different random parameter locations does
not affect the determination of the SFH.

\begin{figure}
\begin{center}
\includegraphics{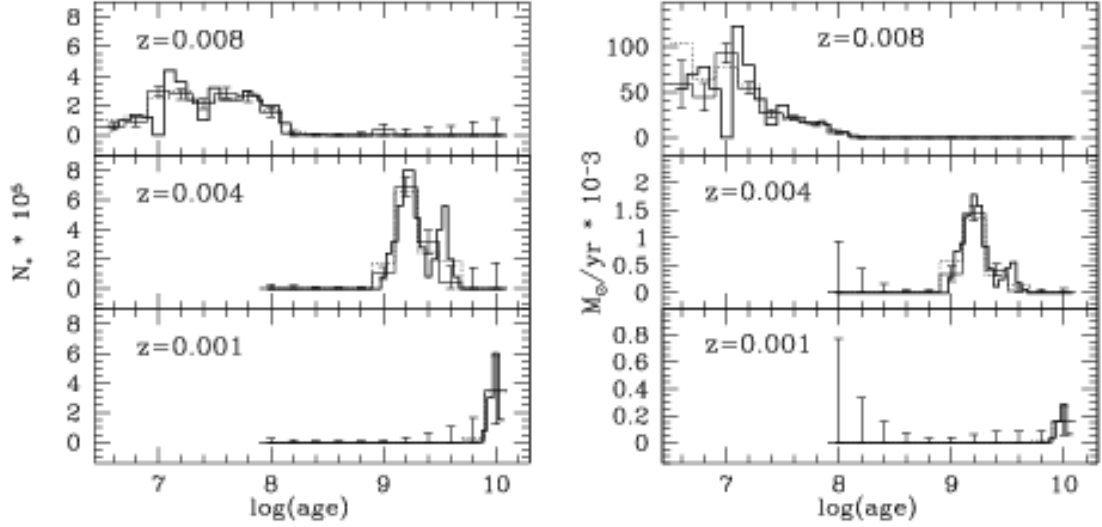}
\end{center}
\figcaption{The recovery of a continuous SFH, in which the SFR varies
within the age range of individual isochrone groups.  The left panels
show the SFH as the number of stars formed per isochrone group.  The
right panels show the SFR, in $M_\sun / yr$.  The heavy solid line is
the input SFH; note the variation of the input SFH within isochrone
groups.  The dotted line indicates the average input SFR over the age
interval of each isochrone group.  The thin solid line with errorbars
is the best-fit recovered SFH.  Note that while the z=0.008 isochrones
form a relatively small number of stars, the SFR for these isochrones
is relatively high, due to the fact that the time axis is logarithmic.  
\label{fig:test2}}
\end{figure}

We simulate more realistic artificial populations by allowing
each isochrone to have an independent SFR.  The fitting algorithm is
inherently unable to fit these populations exactly, because the
synthetic CMDs are each composed of multiple neighboring isochrones
whose amplitudes are set to be equal to reduce the number of
independent isochrones.  Figure \ref{fig:test2} shows that the 
algorithm is able to disentangle the mix of many different stellar 
populations, and it correctly selects the mean SFR over each isochrone 
group.  Note that the simulated SFH shown in Figure \ref{fig:test2} has 
a gap in star formation from 3 to 8 Gyr ago, similar to an age gap
suspected in the LMC.  This gap is reliably recovered by the
algorithm, suggesting that we will be able to answer important
questions about the ancient SFH of the Magellanic Clouds. 

\begin{figure}
\begin{center}
\includegraphics{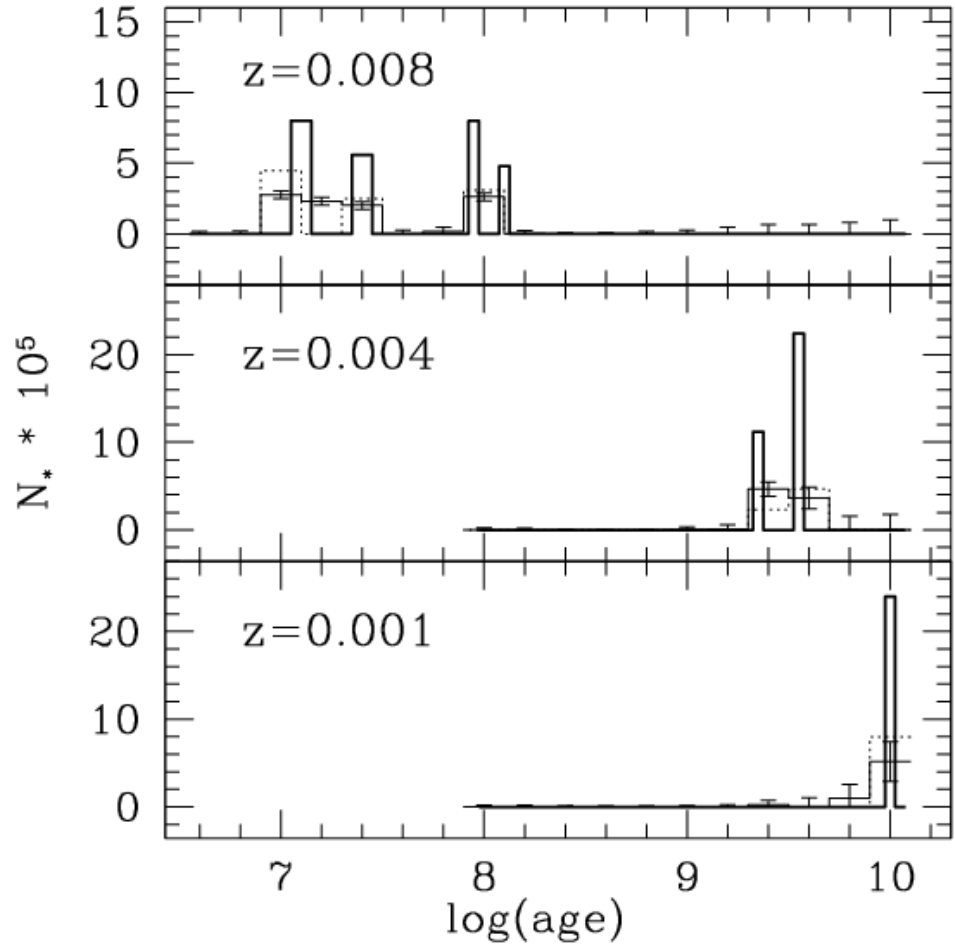}
\end{center}
\figcaption{The recovery of a SFH composed of discrete, instantaneous
bursts.  The heavy solid line shows the input SFH, the dotted line
shows the input SFH, averaged over the isochrone groups, and the thin
solid line with error bars shows the best-fit recovered SFH.
\label{fig:test3}}
\end{figure}

A more challenging test of the SFH finder is that of an instantaneous
burst of star formation, especially one that occurs near the edge of
an isochrone group.  We construct a stellar population based on
several such instantaneous bursts, and recover the best-fit SFH.
Figure \ref{fig:test3} shows that the recovered SFH again consists of
the mean SFR over each isochrone group.  The SFH algorithm gives no 
indication of whether there are variations in the SFR within 
isochrone groups, even when these are as extreme as an instantaneous
burst.  If high-frequency variations in the SFR need to be recovered,
the isochrone groups need to cover smaller ranges in age (\cf\
\S\ref{sec:gc}).  

\subsubsection{Varying model parameters}\label{sec:tpvary}

The SFH extraction requires knowledge of the following external 
parameters:  the distance modulus, the interstellar extinction, the
IMF slope, the photometric errors, and the binary fraction.  We next
investigate the effects on the extracted SFH of having incorrectly
determined these parameters.  

\begin{figure}
\begin{center}
\includegraphics{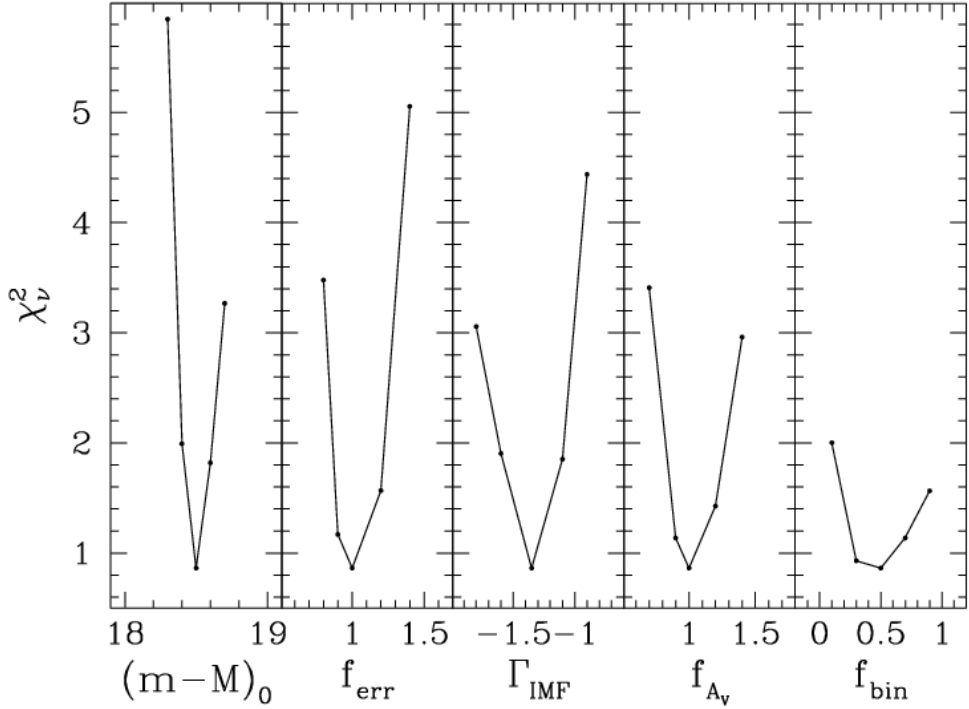}
\end{center}
\figcaption{The effect of varying the external parameter's values on
the recovered SFH.  For each external parameter, we construct a series
of input populations with different parameter values.  In recovering
the SFH of these populations, however, we assume the standard
parameter value appropriate for the LMC.  The panels show the
resulting reduced \cnu\ values for each tested parameter.  In each
case, the best match is found when the correct parameter value is
used, but the fit generally degrades slowly as the parameter value
changes. \label{fig:pvtests}}
\end{figure}

For each external parameter, we construct a series of artificial
stellar populations using the same input SFH from Figure
\ref{fig:test2}, varying the parameter's value for each population.
However, the SFH for each population is determined assuming the
canonical values for these parameters as described in
\S\ref{sec:tpmatch}.  The results of these tests are shown in Figure
\ref{fig:pvtests}.  In each case, the lowest \cnu\ values are found
when the parameter value is correctly determined, suggesting that our
method can be used to constrain these external parameters.  However,
the constraint provided by varying these parameters in our models is
not generally more precise than is available by other methods.

\begin{figure}
\begin{center}
\includegraphics{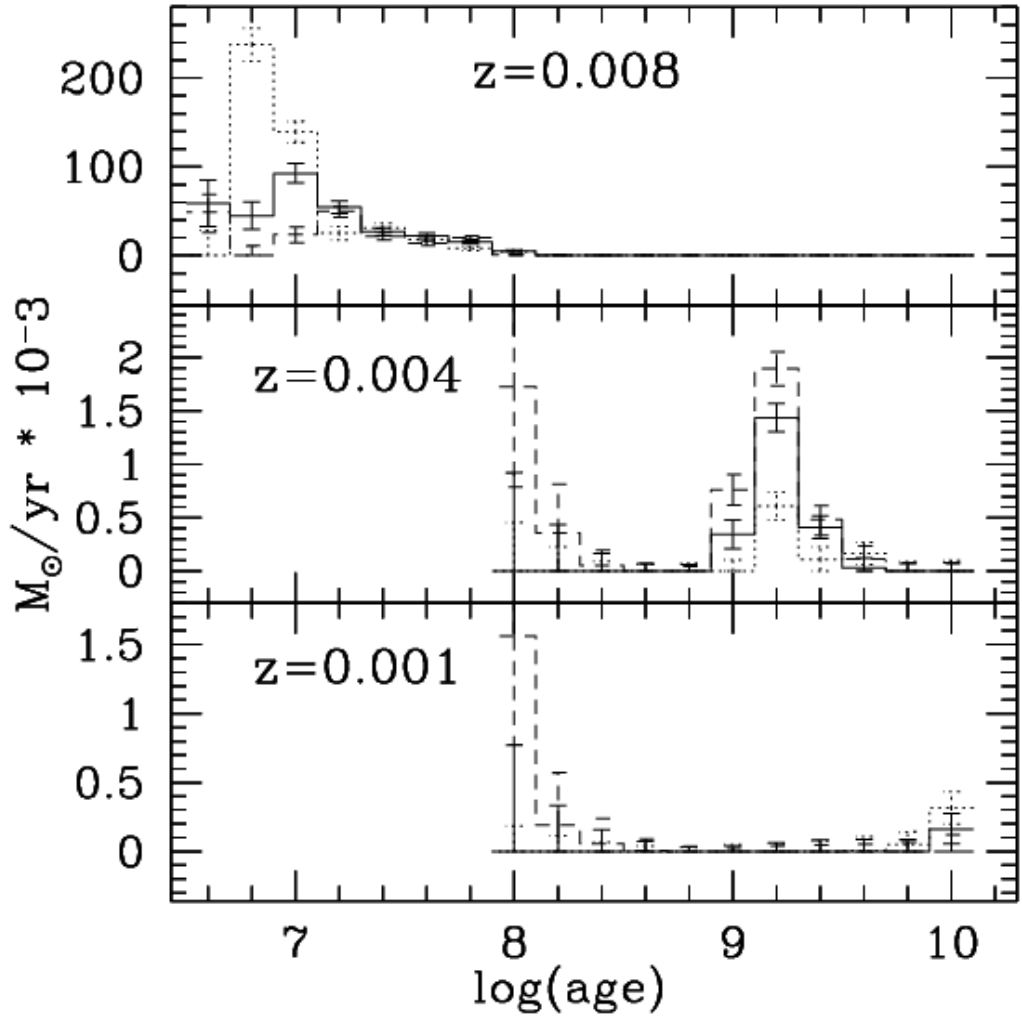}
\end{center}
\figcaption{The recovered SFHs of three artificial populations, 
generated with different values for the IMF slope: $-1.8$ 
(dashed line), $-1.35$ (solid line) and $-0.9$ (dotted line). 
In each case, the recovery algorithm used $\Gamma=-1.35$.
\label{fig:imf}}
\end{figure}

Some of the parameters remain poorly constrained using either our
method or external analyses (notably the
IMF slope and binary fraction).  The IMF slope in particular is often
considered a significant obstacle for synthetic CMD fitting
techniques, because it has a large range of plausible values, and the
value chosen affects the derived SFH significantly.  We examine this
sensitivity by plotting the derived SFHs for the same artificial
population, assuming three different IMF slopes: $\Gamma=-0.9, -1.35$
and $-1.8$ (\cf\ Figure \ref{fig:imf}).  Although there are some
notable systematic differences, the general form of the
derived SFH is remarkably unchanged, considering the wide variance in
IMF slope.  Differences in IMF slope will have a more significant
impact on the SFHs of populations which contain larger numbers of
recently formed massive stars.  In these cases, we recommend
recomputing the SFH with different IMF slope values as we have done. 
While a strong constraint on the true IMF slope may not be possible,
it will at least be clear how sensitive the results are to IMF slope.

\subsection{Globular Cluster Photometry}\label{sec:gc}

The artificial population tests demonstrate the ability of our
algorithm to reliably disentangle the photometry of different stellar
populations, and derive an accurate SFH.  However, in these tests, it
is guaranteed that the synthetic CMDs represent the tested population
reasonably well, even in cases where we impose systematic errors
(\cf\ \S\ref{sec:tpvary}).  In applying the method to real
populations, this guarantee does not exist.  We do our best to
accurately account for the distance, interstellar extinction, IMF, and
photometric errors, but the only way to rate our accuracy (and that of
the theoretical isochrones) is to test the method on real stellar
populations whose SFHs have been measured by isochrone-fitting techniques.  
Globular clusters provide an ideal test case, because they have simple, 
delta-function SFHs and many have well-established age estimates.
We are aware that, like any method that relies on fitting isochrones to 
observed photometry, there is always a possibility that the models contain
unknown systematic errors.  Therefore, even if we get the same SFHs for 
these populations, we have not ``proven'' their age any more than 
previous authors have.  We are merely attempting to establish that our 
method gives consistent results with the more traditional isochrone 
fitting methods.  The method was designed so that, as improved theoretical 
isochrones are published, they can easily be adopted.

\begin{figure}
\begin{center}
\includegraphics{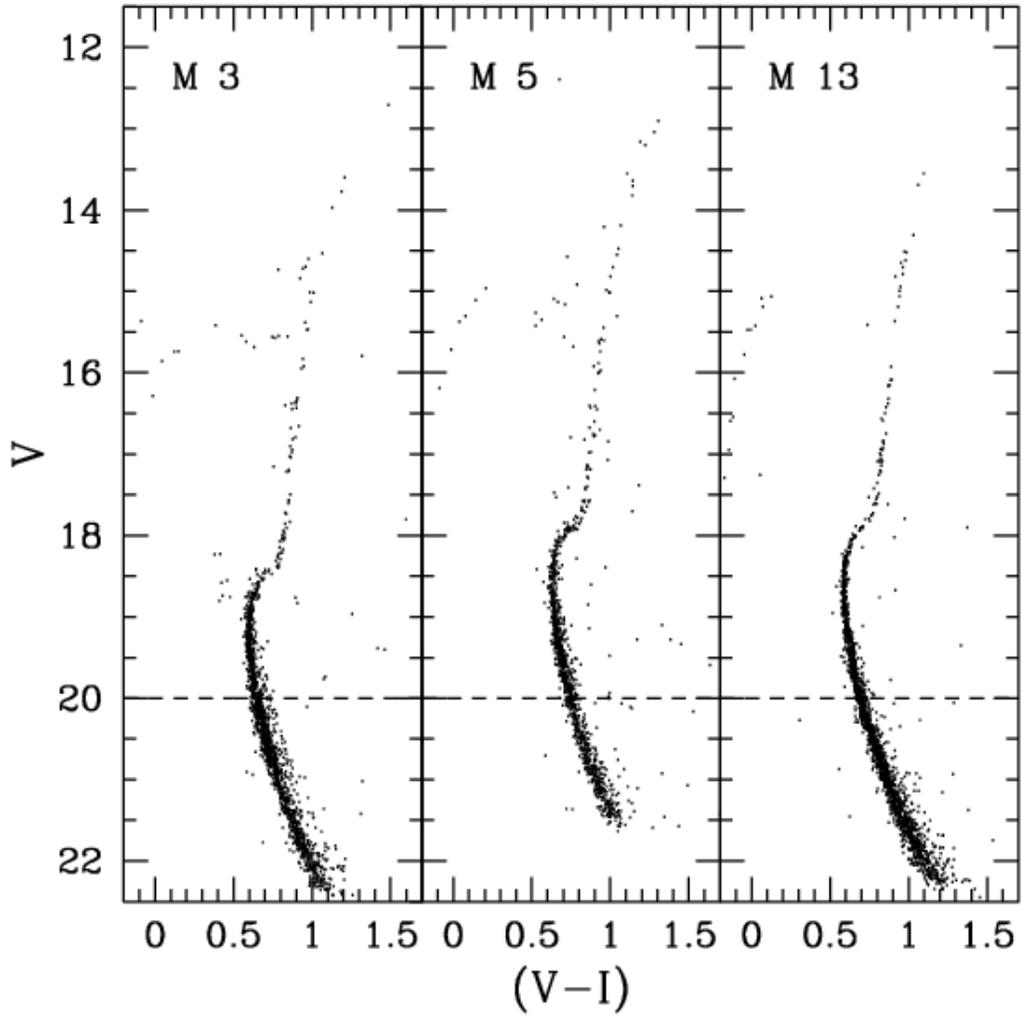}
\end{center}
\figcaption{Deep $\vi$ vs. $V$ CMDs for three Galactic globular
clusters, from \cite{jb98}.  The dashed line represents our
artificial faint limit, imposed to avoid significant completeness
effects.\label{fig:gccmd}}
\end{figure}

We determine the SFH of three Milky Way globular clusters (M 3, M 5,
and M 13) for which deep $V$,$I$ photometry is available \citep{jb98}.  
CMDs representing this photometry are shown in Figure \ref{fig:gccmd}.  
These data are substantially different than the MC Survey data for which 
we have designed the method.
(1) The photometry is complete well below the ancient main
sequence turn off.
(2) The photometric errors are small, eliminating
the need to combine isochrones into wider age bins.
(3) Galactic globular clusters are composed exclusively of old stars.
(4) Globular clusters generally do not suffer from differential
interstellar extinction. 
(5) These data are not crowding-limited, so artificial star tests
are not absolutely necessary for a reasonable characterization of the
photometric errors.
(6) The clusters were imaged in only 2 filters ($V$ and $I$), so only
one CMD can be constructed. 

Our method is easily modified to be appropriate for these data,
despite the many differences from our MC Survey data:  
(1) we modify the time resolution and age range of the synthetic CMD
library, covering ages 3 Gyr to 18 Gyr with a resolution of $\Delta
\log(t)=0.05$;  
(2) we employ an extremely simple photometric error
model, assuming constant photometric errors of $\sigma_V=0.01$ mag and
$\sigma_{\vi}=0.02$ mag, based on visual inspection of the CMDs  (by
using such an {\it ad hoc} error model, we continue to stress-test  
the SFH algorithm by subjecting it to less than ideal conditions); 
(3) because we have not performed artificial star tests on these data,
we avoid the complication of incompleteness by limiting our analysis
to stars brighter than $V=20$ mag (dashed line in Figure
\ref{fig:gccmd});  
(4) we adopt simple foreground-only extinction corrections and
distance moduli, taken from previous studies of these clusters; 
(5) we construct the $\vi$, $V$ CMD only; and
(6) we divide the CMD using a finer grid, to take advantage of the
smaller photometric errors provided by these data.  These changes to the
method's functionality were made without having to recompile the code; all 
operational parameters can be manipulated through input files.

\begin{figure}
\begin{center}
\includegraphics{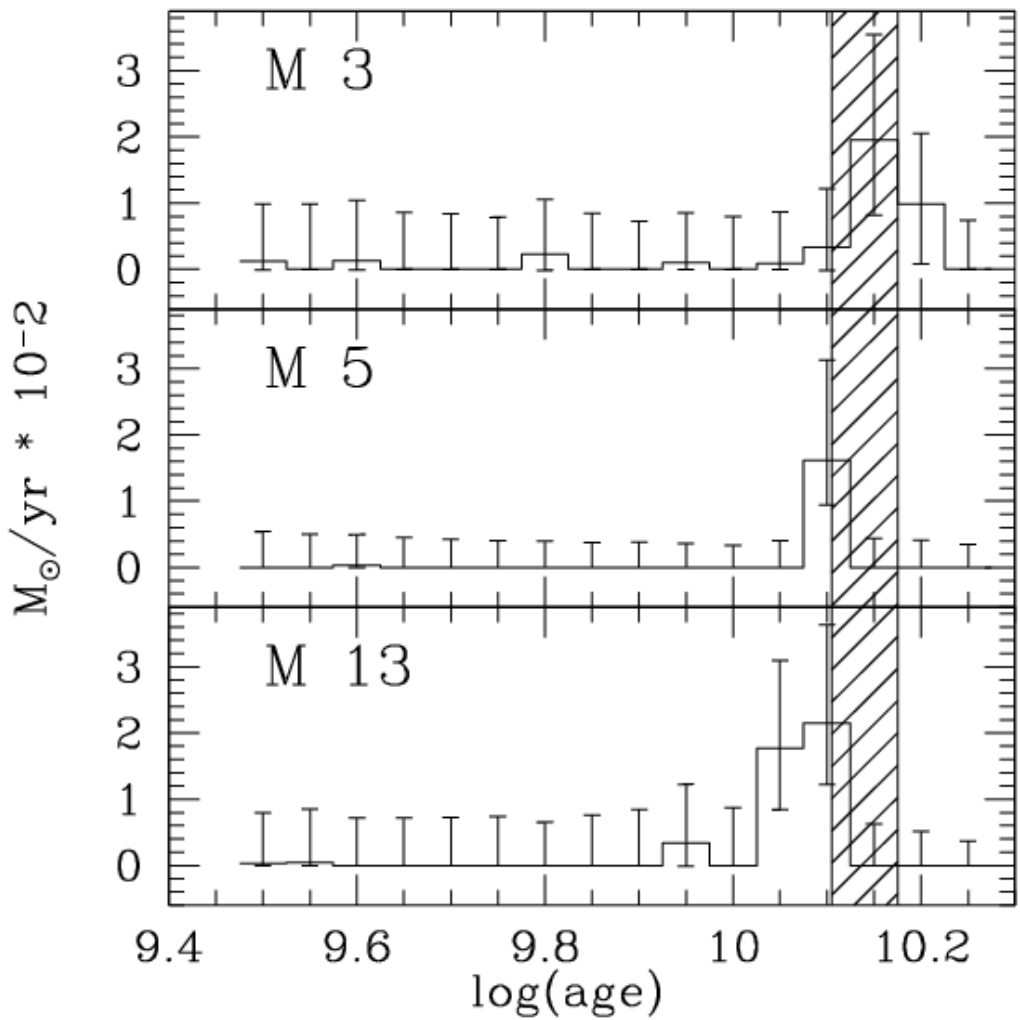}
\end{center}
\figcaption{The recovered SFHs of the three Galactic globular
clusters.  The shaded region indicates the $14\pm1$ Gyr age for these
clusters, as derived by direct isochrone fitting \citep[\cf\ ][and
references therein]{jb98}.  The SFH algorithm was easily modified to
be suitable for these data (\cf\ \S\ref{sec:gc}).  While three
metallicities were included in the fit, only the z=0.001 amplitudes
are shown.  The star formation rate was consistent with zero for the
more metal-rich amplitudes. \label{fig:gcsfh}} 
\end{figure}

Once these changes are implemented, the SFH algorithm selects the SFHs
shown in Figure \ref{fig:gcsfh}.  These three clusters each have
previously published ages of 14 Gyr and metallicities of order
z=0.001 \citep{jb98}.  Figure \ref{fig:gcsfh} shows only the z=0.001
amplitudes.  The star formation rates for isochones at z=0.004 and
z=0.008 were found to be consistent with zero in each cluster.  We
conclude that the algorithm has determined the SFH of these clusters
quite well ($\Delta t/t = 0.1$), considering the crude photometric
error model used, and the fact that Johnson \& Bolte used a different
isochrone set than we used. 

\subsection{MC Survey Photometry}\label{sec:mcsurvey}

The algorithm described in this paper was developed to extract the
star formation history of the Large and Small Magellanic Clouds from
the $UBVI$ photometry of our Magellanic Clouds survey.  We are 
currently reducing and analyzing these data, and the full spatially
resolved SFH of the Clouds will be presented in future papers in this
series.  As a preview of this work, and also to provide an
additional test of our algorithm, we present the SFH extraction of
selected subsets of our MC survey data.  

\subsubsection{NGC 1978}\label{sec:n1978}

\begin{figure}
\begin{center}
\includegraphics{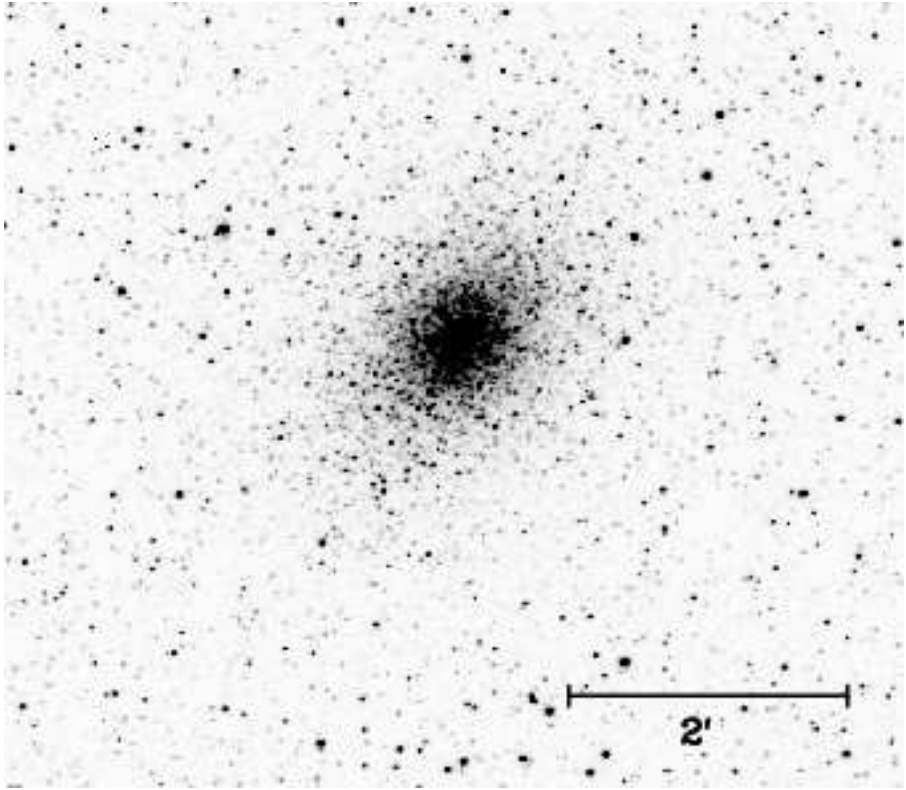}
\end{center}
\figcaption{A $V$-band image of a 7.2 arcmin $\times$ 7.2 arcmin
subsection of our MC Survey, centered on the populous LMC cluster NGC
1978.  These stars provide an ideal test for our SFH algorithm,
because several determinations of the age and metallicity of NGC 1978
exist in the literature. \label{fig:n1978img}} 
\end{figure}

\begin{figure}
\begin{center}
\includegraphics{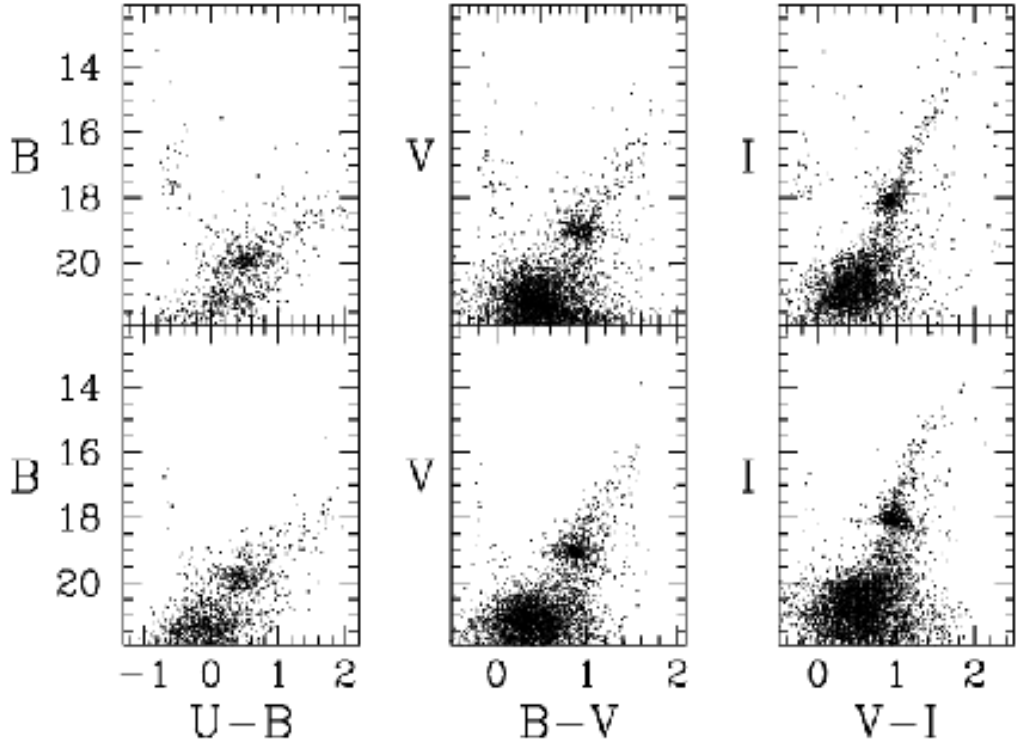}
\end{center}
\figcaption{The color-magnitude diagram triplet for NGC 1978, from our
MC Survey photometry.  Contaminating field stars have been removed from
these data through the statistical subtraction of stars in a nearby
field of the same size.  However, some residual field contamination
is visible in the upper main sequence region.  Note that the algorithm
has attempted to fit this contamination.\label{fig:n1978cmd}} 
\end{figure}

\begin{figure}
\begin{center}
\includegraphics{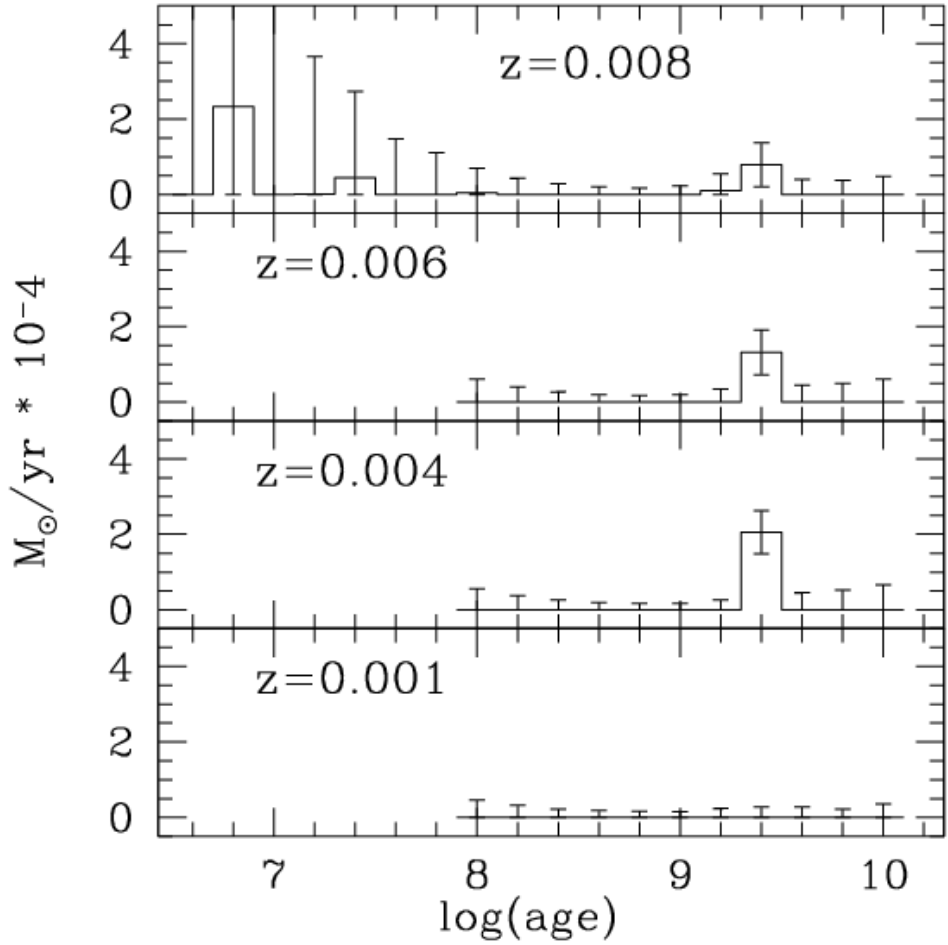}
\end{center}
\figcaption{The best-fit recovered SFH for NGC 1978, showing a
dominant burst at 2.5 Gyr, with a significant metallicity spread.
We compare this result with previous age and metallicity measurements
of NGC 1978 in Table \ref{tab:n1978}. \label{fig:n1978sfh}}
\end{figure}

We first present the SFH of the populous star cluster NGC 1978.  This
is a well-studied star cluster, with numerous age and metallicity
determinations in the literature (\cf\ Table \ref{tab:n1978}).  It
therefore provides an ideal introduction to our MC Survey photometry
because we can check our findings against these previous determinations.
We select a rectangular subregion of our Survey, centered on NGC 1978,
and spanning 7.2 arcmin in right ascension and declination (\cf\ Figure
\ref{fig:n1978img}).  The stars within an arcminute of the cluster center 
are extremely crowded in our data.  Because these stars have unacceptably
large photometric errors, we exclude them from our sample.  We also
perform a statistical subtraction of background field stars, using a
nearby control field of the same angular size.  The statistically
cleaned NGC 1978 CMDs are shown in the top panels of Figure
\ref{fig:n1978cmd}, and the best-fit SFH for this stellar population
is shown in Figure \ref{fig:n1978sfh}.  We find a dominant burst with
an age of 2.5 Gyr ($\log(t) = 9.4$), and a metallicity spanning $-0.7
< [Fe/H] < -0.4$.  CMDs for the artificial population constructed from
this SFH are shown in the bottom panels of Figure \ref{fig:n1978cmd}. 
The wide range in metallicities found is likely an indication that the
method is unable to resolve metallicity differences of $\simless 0.3$
dex in such data.  Synthetic CMDs with z=0.006 are photometrically too
similar to both z=0.004 and z=0.008 synthetic CMDs for our method to
distinguish between them, given the current data.  Therefore, we
exclude synthetic CMDs with z=0.006 from our library.  Regardless of
this issue, the age and metallicity found for NGC 1978 are consistent
with the results from previous studies (\cf\ Table \ref{tab:n1978}),
demonstrating that the method has correctly determined the SFH of this
population.  

\subsubsection{LMC field stars}

\begin{figure}
\begin{center}
\includegraphics{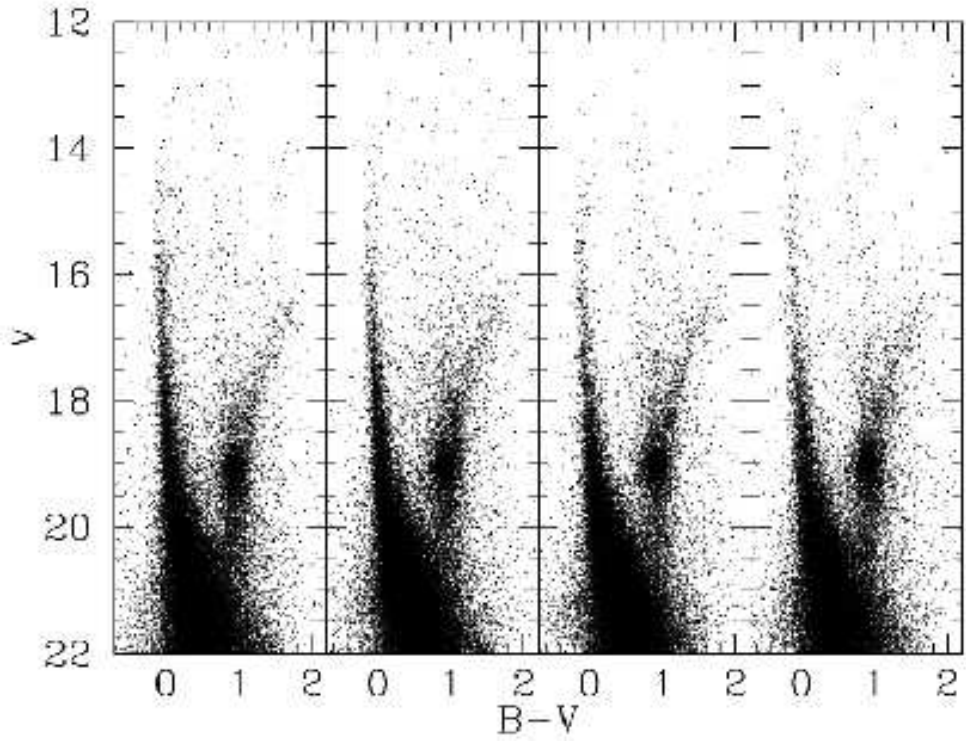}
\end{center}
\figcaption{$\bv$ vs. $V$ CMDs of four 20 arcmin $\times$ 20 arcmin
subregions of our MC Survey photometry.  These fields are adjacent
in right ascension.  They are north of the LMC bar, centered at
approximately ($5^h \: 27m, -66\degrees \: 6'$).  The differences
in the morphologies of these CMDs hint that variations in the SFH are 
significant on this scale.  \label{fig:lmccmd}}
\end{figure}

As a final demonstration of the algorithm, we determine the SFHs of 
four contiguous 20 arcmin $\times$ 20 arcmin regions in the northern
LMC.  The four regions are adjacent in RA, and cover 
$5^h \: 21^m < \alpha < 5^h \: 33^m$ and 
$-66\degrees \: 16' < \delta < -65\degrees \: 56'$.  
Our MC Survey catalog contains $UBVI$ photometry for about 50000
stars in each of these regions (\cf\ Figure \ref{fig:lmccmd}).  
The full spatially-resolved SFH of both the SMC and LMC, consisting of
the individual SFHs of hundreds of regions of this size, will be 
presented in future papers in this series.

\begin{figure}
\begin{center}
\includegraphics{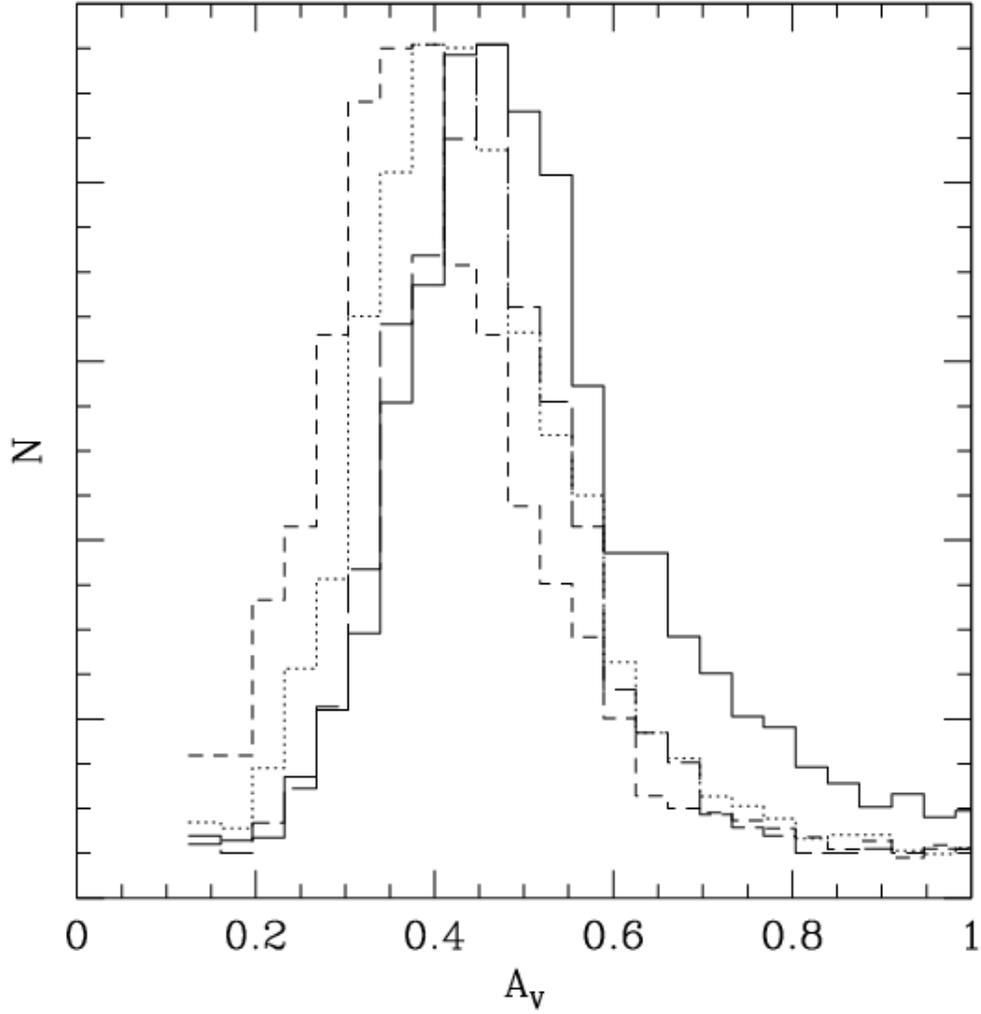}
\end{center}
\figcaption{A comparison of the distribution of extinction values 
toward hot stars in each of the four LMC regions.  Each distribution 
represents our measured extinction values in one of the four adjacent 
LMC regions, arbitrarily normalized.  The extinction varies significantly 
between these regions, so we construct a unique set of synthetic CMDs 
for each region, each using its own local extinction distribution.
\label{fig:lmcext}}
\end{figure}

We construct an independent synthetic CMD library for each region.  
We use a single artificial star test for all four regions, since each
region contains approximately the same density of stars.  However,
because the reddening varies significantly in these regions (\cf\
Figure \ref{fig:lmcext}), each region uses its own local reddening
distribution in constructing the synthetic CMDs.  We stress the
importance of using an accurate reddening model.  Our first attempt at
recovering the SFHs of these regions utilized a single reddening
distribution.  The resulting SFHs had \cnu\ values that were larger
than the current values by a factor of two.

\begin{figure}
\begin{center}
\includegraphics{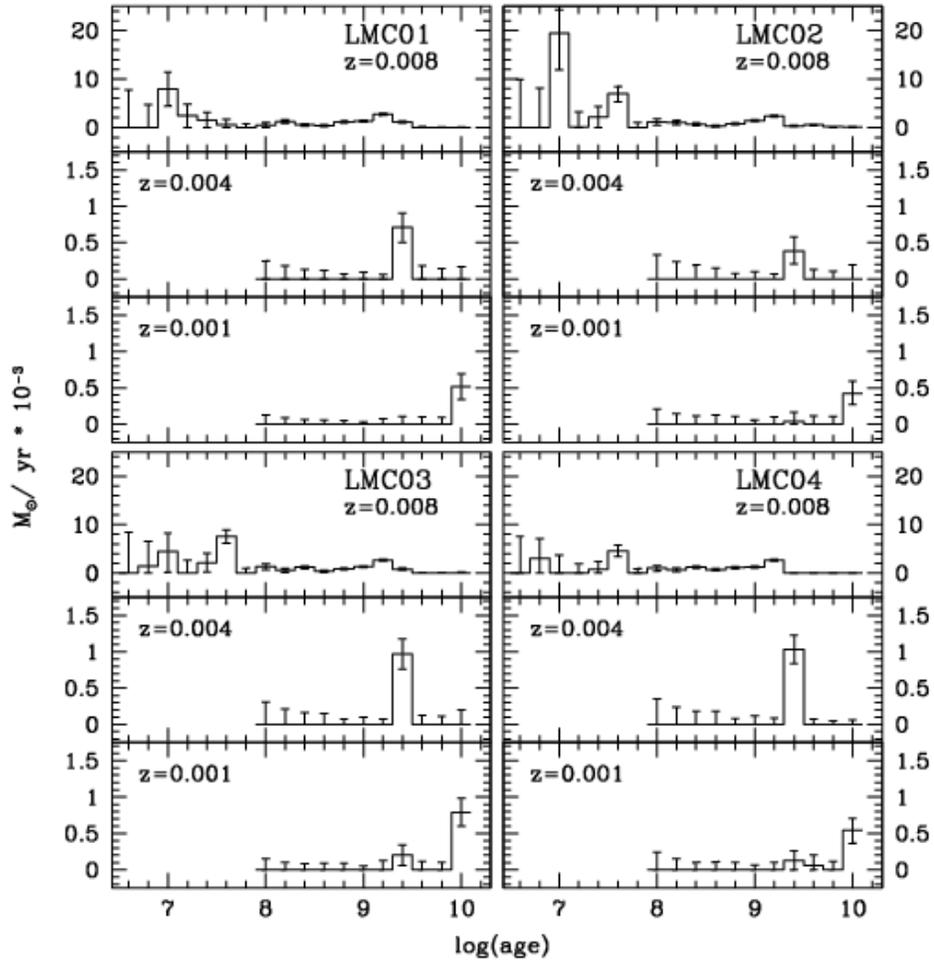}
\end{center}
\figcaption{The best-fit recovered SFHs for the four subregions shown
in Figure \ref{fig:lmccmd}.  \label{fig:lmcsfh}}
\end{figure}

Once the synthetic CMD libraries are constructed, the algorithm
selects the best-fit SFHs shown in Figure \ref{fig:lmcsfh}.  For 
each model, the reduced \cnu\ is between 7.0 and 9.0; most of the difference 
is due to a small apparent color offset along the main sequence, in the sense
that the models are slightly bluer than the data.  The derived  
SFHs are generally consistent with the recent reconstruction of the 
LMC SFH by \citet{hol99}; we both find a burst of star formation 2-4 
Gyr ago, followed by a lower, continuous star formation rate, up to 
the present day.  

There are several interesting features to note in Figure
\ref{fig:lmcsfh}.  The recent SFH is highly variable, and changes
significantly between regions.  There are peaks in the recent star
formation rate at 10 and 40 Myr, and the relative strengths of these
peaks change systematically from region to region.  The SFHs older
than $\sim1$ Gyr are nearly indistinguishable, as expected due to
dynamical mixing of stellar populations.  The old SFH is dominated by
two bursts; a z=0.001 burst at 10 Gyr, and a z=0.004 burst at 2.5 Gyr.
This intermediate-age burst has the same age as NGC 1978, which is
contained in LMC03.  It is possible that these stars are dissolved NGC
1978 members, now populating the LMC field up to 500 pc away from the
cluster.  

We observe a consistent age-metallicity relation in these regions. 
The oldest stars have z=0.001, and there is no significant population
at this metallicity younger than 2.5 Gyr.  This star formation is
followed by the single burst with z=0.004, 2.5 Gyr ago.  All
populations younger than this burst have z=0.008, suggesting a rapid
increase in metallicity following the burst.  This observed
age-metallicity relation is consistent with the analytic  
model of \citet{pt98}.

\begin{figure}
\begin{center}
\includegraphics{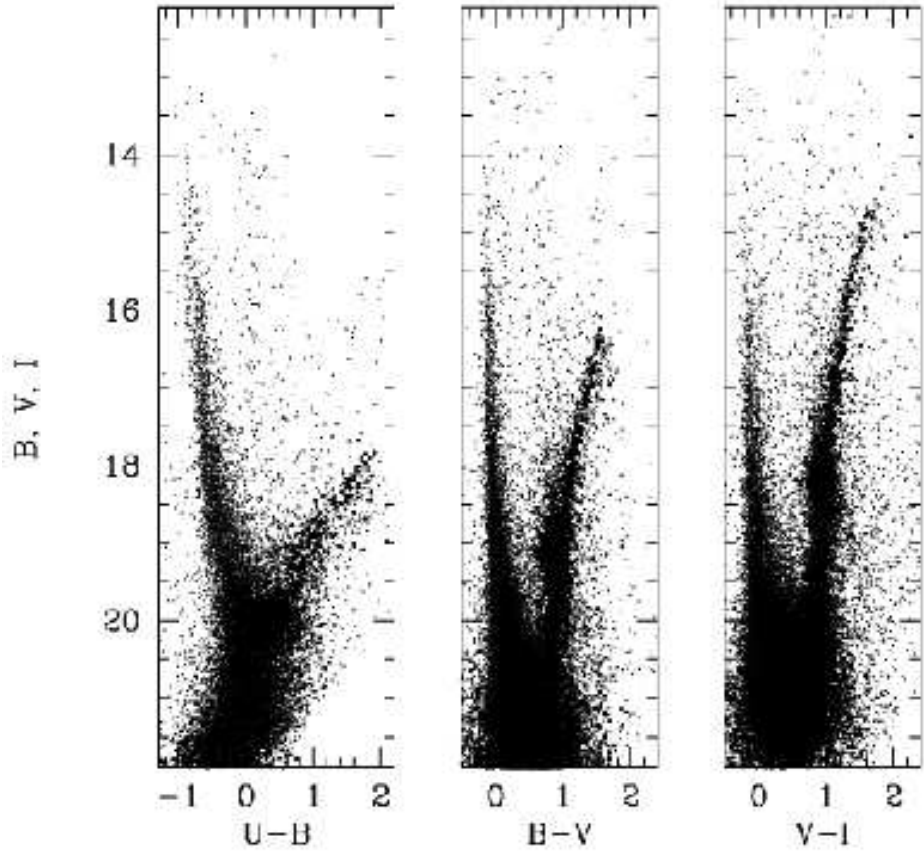}
\end{center}
\figcaption{CMD triplet showing the photometry of one of our LMC regions
(small points) with a supplemental artificial population (large points) 
added to fill the age gap between 3 and 8 Gyr. \label{fig:nogapcmd}}
\end{figure}

\begin{figure}
\begin{center}
\includegraphics{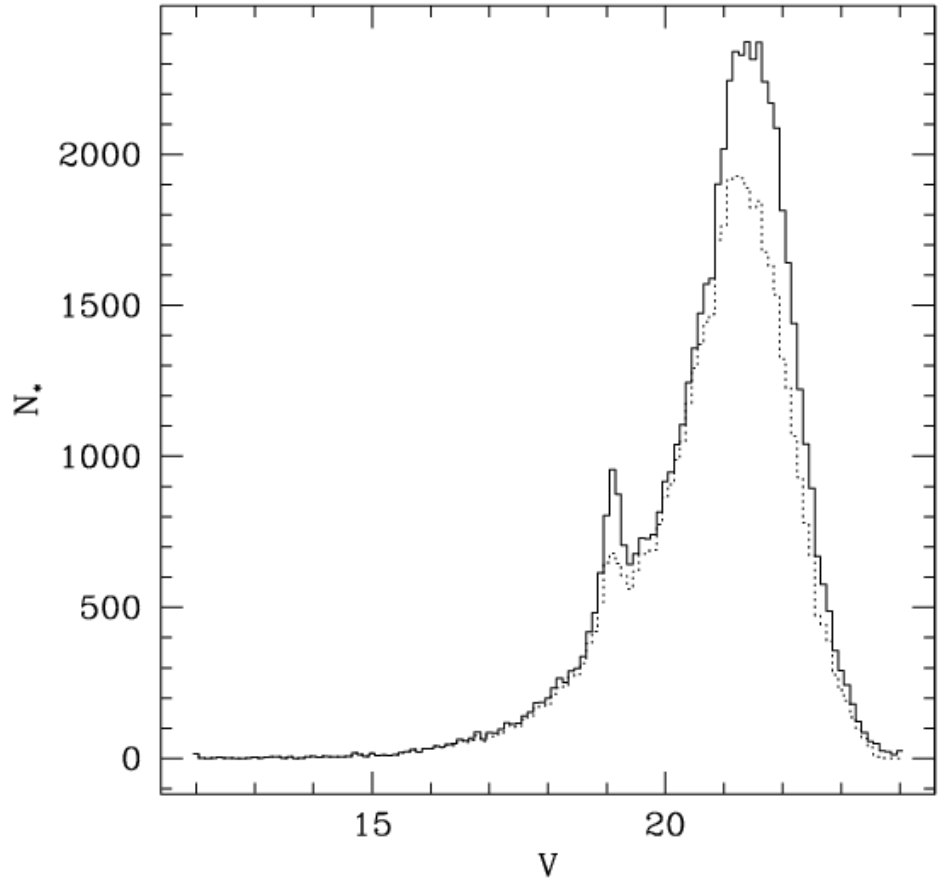}
\end{center}
\figcaption{The luminosity function of the LMC region shown 
in Figure \ref{fig:nogapcmd}, both with (solid line), and without 
(dotted line) the supplemental population. \label{fig:nogaplf}}
\end{figure}

\begin{figure}
\begin{center}
\includegraphics{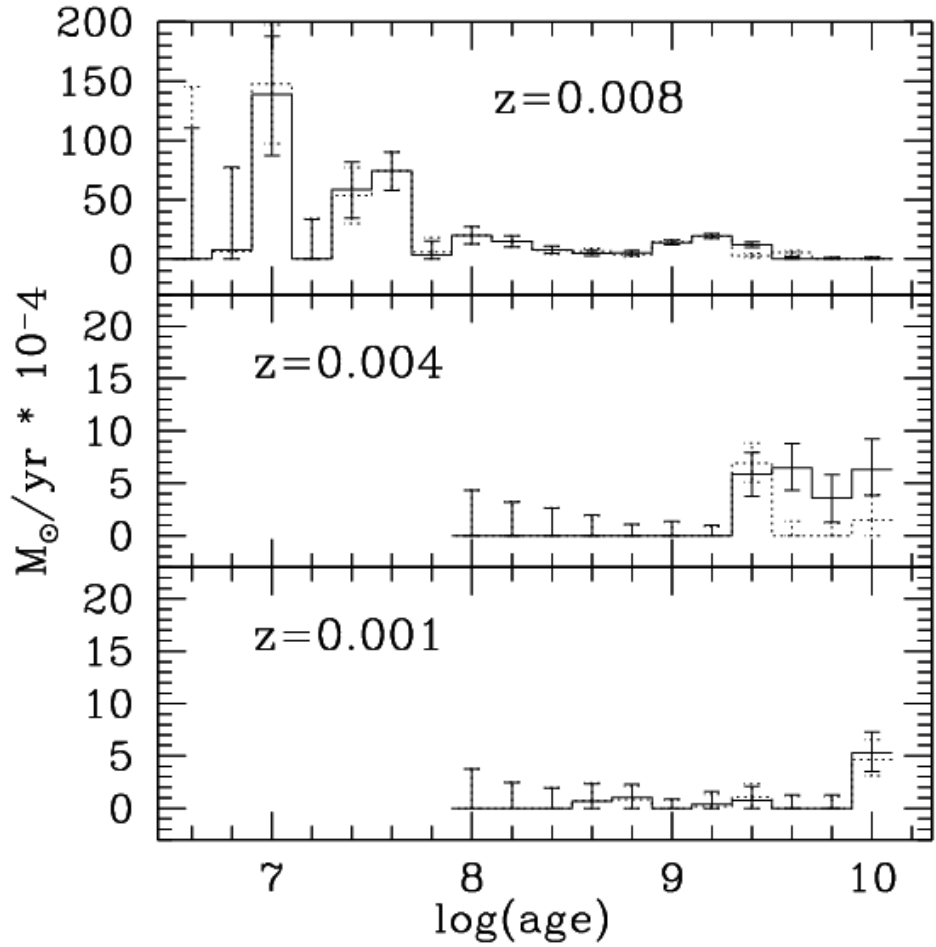}
\end{center}
\figcaption{The best-fit recovered SFH of the LMC region shown in Figure
\ref{fig:nogapcmd}, both with (solid lines) and without (dotted lines)
the supplemental population, added such that the total star formation rate
is constant between 2 and 10 Gyr. \label{fig:nogapsfh}}
\end{figure}

Finally, we note that there is no evidence for significant star 
formation between 3 and 8 Gyr, in agreement with the well-known gap  
in the ages of LMC clusters.  This is a tantilizing result that
warrants a closer look.  The amplitude error bars suggest a high
degree of confidence that the gap exists, yet the gap is not obvious
in the CMDs of these populations (Figure \ref{fig:lmccmd}).  What is
driving the algorithm to this solution?  This question is difficult to
address directly, because the algorithm is designed to consider the
entire multicolor photometric distribution simultaneously.  One could
restrict the fit to certain portions of the CMDs, but rather than attempt 
to second-guess the algorithm, we decided to instead ask the question:
how would the CMDs look if the gap did not exist?  In Figure
\ref{fig:nogapcmd}, we present the observed photometry from one of the LMC 
regions, with a supplemental, artificial population added to fill the age gap. 
The size of the population is such that the effective SFR remains
approximately constant over the interval from 10 Gyr to 2 Gyr ago.  
The new population's metallicity is z=0.004.  The supplemental population
occupies the red giant branch and faint main sequence regions, but it is not
obviously distinct from existing populations in the field.  Despite this, the
algorithm recovers the correct SFH when given this composite population 
(\cf\ Figure \ref{fig:nogapsfh}).  The algorithm is able to use small
distinctions, such as the position of the red giant branch and the number of 
main sequence stars as a function of magnitude, to infer the presence of the 
supplemental population.  Because red giant branch morphology is relatively 
insensitive to age, the main sequence stars likely dominate the determination 
of ages in the SFH.  The supplemental population looks quite subtle in Figure
\ref{fig:nogapcmd}.  Its presence is more easily seen in Figure
\ref{fig:nogaplf}, which compares the luminosity functions of the original 
population and the composite population.  The significance of faint
main-sequence stars in determining the old SFH underscores the importance of 
accurately estimating photometric errors, completeness rates and
the interstellar extinction.  Each of these parameters has an important effect
on the relative number of faint main sequence stars in the synthetic CMDs.

\section{Impact of Systematic Isochrone Errors}\label{syserrs}

The accuracy of theoretical models is a fundamental issue in attempting to 
reconstruct star formation histories.  While we can be confident that some 
linear combination of theoretical isochrones can accurately reproduce our 
observed photometry (\cf\ Figure \ref{fig:n1978cmd}), we still cannot be sure 
that the {\em correct} combination of isochrones produces the best fit.  
This was our primary motivation in ensuring that any set of isochrones 
can be used with the method; as improved isochrones are published, they can
easily be inserted.  Also, because we resolve the CMDs rather coarsely (we 
use 0.25 mag $\times$ 0.25 mag grid boxes), we do not require absolute 
precision and accuracy in the isochrones.  Other than provide the  
flexibility to accept new models and robustness against small errors, there 
is little that can be done to defend against such systematic errors, if they 
exist.  Even if we are reasonably confident that the isochrones are positioned
correctly in CMDs (thanks to decades of work comparing isochrone tracks to 
cluster photometry), a more subtle issue remains: whether the models predict 
the correct rate at which stars age along evolutionary tracks.  If the
evolutionary rates are incorrect, the ``occupation probabilities'' of the 
isochrones will also be wrong.  This possibility is especially troubling for 
our ancient SFHs: if the RGB region is systematically underpopulated by 25\%, 
then our 10-Gyr amplitude could be just a compensation for the ``missing'' 
RGB stars from intermediate-age populations (note, however, that the RGB
morphologies of 10-Gyr and 2.5-Gyr populations are not quite the same, which 
may limit the ability of the ancient population to substitute for the
intermediate-age RGB).  Such an error could be ruled out by comparing the 
luminosity function of a single-burst stellar population to that of a 
synthetic population produced by the best-fit isochrone to the real 
population.  Such a comparison was recently performed by \cite{zp00}.  They 
compared the luminosity functions of 18 globular clusters, based on HST
photometry, to the predicted luminosity functions from 4 sets of isochrones
(including the Padua isochrones that we employ).  They conclude that the
synthetic luminosity functions match those of the globular clusters, to within
HST's photometric errors.

\section{Summary}\label{sec:summ}

We have presented a method for determining the star formation history
of resolved stellar populations.  By adding the effects of distance,
interstellar extinction, the initial mass function, photometric
errors, and binarity to theoretical isochrones, we construct a library
of template synthetic CMDs, each representing the photometric
distribution of a stellar population with a single age and
metallicity.  These synthetic CMDs are then combined linearly and
compared quantitatively to observed stellar photometry, to determine
the star formation rate as a function of age and metallicity.  In this
way, present-day stellar populations can be used as a fossil record of
a galaxy's remote history.   

Using both artificial photometry and real data, we have extensively
tested the robustness and accuracy of the method, under a
variety of conditions.  We find that the method selects the correct
star formation history, even when the external parameters (distance modulus, 
IMF slope, binary fraction, interstellar extinction, and photometric errors) 
are imperfectly determined.  We also apply the method to four fields in the 
Large Magellanic Cloud.  We find significant spatial variation in the recent 
star formation history, and a consistent age-metallicity relation in all four
fields.  We also observe a quiescent era between 3 and 8 Gyr 
ago, similar to the well-known age gap seen among LMC star clusters.

We will apply the method to the photometry from our survey of the
Magellanic Clouds.  We have obtained $UBVI$ photometry 
of tens of millions of stars in each Cloud; for the first time, we
have a complete census of the stellar populations down to $V\sim21$
mag in these two galaxies.  By dividing these data into hundreds of
subregions, we will construct a map of the star formation history in
each of the Clouds.  As a demonstration of this work, we present in
this paper the star formation histories for four adjacent regions in
the northern LMC.  These four regions already show significant
variation in their star formation histories, and confirm the presence
of an extended quiescent era in the LMC's remote past.  The full maps
of the SFHs of the Magellanic Clouds will allow us to address many of
the outstanding problems in galaxy evolution and the physical
processes governing star formation.

The method is useful beyond our immediate plans to map out the SFH of
the Magellanic Clouds.  The Hubble Space Telescope and large
ground-based telescopes have the ability to resolve more distant local
group galaxies into individual stars.  Galaxies in the local group
span a wide range in properties, and recent work has suggested that
these galaxies have experienced quite different star formation
histories.  Our method provides a way to uniformly and quantitatively
determine the star formation histories of these systems, so that their
histories can be reliably compared, and explanations of the
differences can be sought.  The algorithm was designed to be easily
adaptable to nearly any kind of stellar photometry data, and the code
is available for the astronomical community at
http://ngala.as.arizona.edu/mcsurvey/SFH 

\vskip 1in
\noindent Acknowledgements:
D. Z. acknowledges financial support from an NSF grant (AST 96-19576), 
a NASA LTSA grant (NAG 5-3501), a David and Lucile Packard Foundation 
Fellowship, and an Alfred P. Sloan Fellowship. J. H. is grateful for
fruitful discussions on synthetic CMD fitting methods with Carme
Gallart and Jon Holtzman. We thank Jennifer Johnson and Michael Bolte 
for making their photometry of M 3, M 5, and M 13 available to us.

\begin{deluxetable}{lcc}
\tablecolumns{3}
\tablewidth{0pt}
\tablecaption{Measurements of NGC 1978 \label{tab:n1978}}
\tablehead{
   \colhead{Author} & \colhead{[Fe/H]} & \colhead{age (Gyr)}
}
\startdata
Present work\tablenotemark{a}    & $-0.4$--$-0.7$  & $2.5\pm0.5$   \\
\cite{bic86}\tablenotemark{d}    & $-1.1$          & 6.6           \\
\cite{bom95}\tablenotemark{b}    & $-0.4$          & 2.1           \\
\cite{chi86}\tablenotemark{b}    & $-0.5$          & 1.8           \\
\cite{dfp98}\tablenotemark{c,e}  & $-0.6$          & $3.0\pm1.0$   \\
\cite{ef88}\tablenotemark{b}     & \nodata         & 2.5           \\
\cite{oz91}\tablenotemark{f}     & $-0.41$         & \nodata       \\
\cite{oz84}\tablenotemark{b}     & $-0.5$          & 2.0           \\
\enddata
\tablenotetext{a}{synthetic CMD fitting}
\tablenotetext{b}{isochrone CMD fitting}
\tablenotetext{c}{integrated broadband photometry}
\tablenotetext{d}{integrated narrow-band photometry}
\tablenotetext{e}{integrated spectrum}
\tablenotetext{f}{stellar spectra}
\end{deluxetable}
\clearpage

\clearpage

\clearpage

\clearpage

\clearpage

\clearpage

\end{document}